\documentclass[letterpaper,prc,twocolumn,showpacs,floatfix,nofootinbib,preprintnumbers,superscriptaddress,amsmath,amssymb]{revtex4-1}

\usepackage{amsmath}           
\usepackage{amsfonts}
\usepackage{graphicx}          
\usepackage{dcolumn}           
\usepackage{bm}
\usepackage{multirow}
\usepackage{array}

\usepackage{color}

\newcommand{\gras}[1]{\boldsymbol{#1}}

\begin{document}

\preprint{\fbox{\sc version of \today}}

\title{Description of Induced Nuclear Fission with Skyrme Energy
Functionals: II. Finite Temperature Effects}

\author{N. Schunck}
\affiliation{Physics Division, Lawrence Livermore National Laboratory, Livermore, CA 94551, USA}

\author{D. Duke}
\affiliation{School of Computing, University of Leeds, UK}

\author{H. Carr}
\affiliation{School of Computing, University of Leeds, UK}

\date{\today}

\begin{abstract}
Understanding the mechanisms of induced nuclear fission for a broad range of
neutron energies could help resolve fundamental science issues, such as the
formation of elements in the universe, but could have also a large impact on
societal applications in energy production or nuclear waste management. The
goal of this paper is to set up the foundations of a microscopic theory to
study the static aspects of induced fission as a function of the excitation
energy of the incident neutron, from thermal to fast neutrons. To account for
the high excitation energy of the compound nucleus, we employ a statistical
approach based on finite-temperature nuclear density functional theory with
Skyrme energy densities, which we benchmark on the $^{239}$Pu(n,f) reaction.
We compute the evolution of the least-energy fission pathway across
multidimensional potential energy surfaces with up to five collective variables
as a function of the nuclear temperature, and predict the evolution of both the
inner and outer fission barriers as a function of the excitation energy of the
compound nucleus. We show that the coupling to the continuum induced by the
finite temperature is negligible in the range of neutron energies relevant for 
many applications of neutron-induced fission. We prove that the concept of 
quantum localization introduced recently can be extended to $T>0$, and we 
apply the method to study the interaction energy and total kinetic energy of 
fission fragments as a function of the temperature for the most probable 
fission. While large uncertainties in theoretical modeling remain, we conclude 
that finite-temperature nuclear density functional may provide a useful 
framework to obtain accurate predictions of fission fragment properties.
\end{abstract}


\maketitle

%
%
%

\section{Introduction}
\label{sec-introduction}

One of the most important challenges for a theory of induced fission is the
capability to predict the evolution of observables such as the charge, mass,
relative yields, total kinetic energy, total excitation energy, and decay
patterns of fission fragments as a function of the energy of the incident
neutron. Recall that the energy of neutrons produced in induced fission follows
roughly a Maxwellian distribution, and the energy range of interest for
applications is typically comprised between a few eV and up to about 14 MeV
\cite{chadwick2011,watt1952}. Following an original idea by Bohr and Wheeler,
induced fission is modeled as the break-up of the compound nucleus formed by
absorption of the incident neutron \cite{bohr1939}. In this picture, neutron
kinetic energies of the order of the MeV correspond to very high excitation
energies of the compound nucleus, where the nuclear level density is very
large \cite{bohr1975}.

In a density functional theory (DFT) approach to induced fission, one may be
tempted to describe such highly excited states directly, via various general
schemes such as the random phase approximation or the generator coordinate
method. However, even assuming all of these methods were properly defined for
the kind of energy densities used in practice (cf. the discussions about
multi-reference density functional theory in
Refs.~\cite{duguet2009,lacroix2009,bender2009,stoitsov2007,anguiano2001-a}), the very large density
of states to consider may jeopardize the success of such a strategy. In
addition, it is expected that dissipation plays a role in fission, and
extensions of these methods to account for explicit couplings between
collective and intrinsic degrees of freedom have only recently been outlined
\cite{bernard2011}.

In this context, the finite-temperature formulation of the nuclear density
functional theory provides an appealing alternative
\cite{eschrig1996,parr1989,blaizot1985,descloizeaux1968}. Assuming that the system is described
by a mixed quantum state uniquely determined by the form of the statistical
density operator provides a convenient basis to quantify the impact of
excitation energy on the deformation properties of the compound nucleus.

There have been many applications of the finite-temperature formalism in
nuclear structure, including early studies of fission barriers using the
Thomas-Fermi approximation
\cite{garcias1990,garcias1989,guet1988,dalili1985,nemeth1985,diebel1981}, the Hartree-Fock
(FT-HF) approximation \cite{bartel1985,brack1974}, and more recently at the
Hartree-Fock-Bogoliubov (FT-HFB) approximation \cite{mcdonnell2013-a,sheikh2009,pei2009},
or applications in the calculation of Giant Dipole Resonances and level
densities \cite{hilaire2012,egido2000,egido1993,egido1988}. Until now, however, there has
been no systematic study of the validity and applicability of
finite-temperature DFT in the description of induced fission. Of particular
importance are the evolution of scission configurations and of fission fragment
properties as a function of the excitation energy of the compound nucleus.

In a previous paper, thereafter referred to as (I), we have used the nuclear
DFT with Skyrme energy densities to analyze static properties of the
neutron-induced fission of the $^{239}$Pu nucleus \cite{schunck2014}. In
particular, we have discussed the role of triaxiality at scission, the
dependence on the parametrization of the energy density functional (EDF) --
including the pairing channel, and the critical importance of scission
configurations. The topological method that we have proposed to identify the
latter allows to define a region in the collective space where scission should
take place. We have then shown that localization techniques borrowed from
electronic structure theory can allow us to approach the asymptotic conditions
of two independent fission fragments. This is key to extracting theoretically
sound estimates of the total excitation energy of the fragments.

Building on this previous study, the goals of this second paper are, therefore,
(i) to establish and validate the framework for nuclear DFT calculations at
finite temperature in the specific context of induced fission, (ii) to study
the evolution of fission barriers and the position and nature of scission
configurations as functions of the excitation energy of the incident neutron,
and (iii) to explore the consequences of the finite-temperature description for
the determination of fission fragment properties. This paper is the second in a
series of several articles focusing on the microscopic description of induced
fission within the framework of the nuclear density functional theory with
Skyrme energy densities.

Section \ref{sec-theory} contains a brief reminder of the theoretical
framework, from basic definitions and concepts related to neutron-induced
nuclear fission to the extension of nuclear density functional theory at finite
temperature with Skyrme functionals. Section \ref{sec-pathway} focuses on the
evolution of potential energy surfaces and fission barriers with temperature.
Section \ref{sec-scission} is devoted to fission fragment properties at finite
temperature, including the extension of the concept of quantum localization, a
study of the coupling to the continuum, and an estimate of the nuclear and
Coulomb interaction energy of the fission fragments.


\section{Theoretical Framework}
\label{sec-theory}

Our theoretical approach is based on the local density approximation of the
energy density functional (EDF) theory of nuclear structure. We recall in the
next few sections some of the basic ingredients of the EDF theory pertaining to
the description of induced nuclear fission at given excitation energy.


\subsection{Thermodynamical and Statistical View of Neutron-Induced Fission}
\label{subsec-thermo}

We begin by recalling a few well-known facts about neutron-induced fission in
order to avoid confusions about the vocabulary used in this work. For fissile
elements such as $^{239}$Pu, the capture of a thermal neutron (in equilibrium
with the environment and with an average kinetic energy of the order of
$E_{n} \approx 0.02$ eV) is sufficient to induce fission. The energy balance 
of the reaction is such that the compound nucleus $(Z,N)$ formed after the 
neutron has been captured is at an excitation energy equal to $|S(N)|$, where 
$S(N)$ is the one-neutron separation energy. In fissile elements, this quantity 
is larger than the fission barrier height, leading to fission. Note that the 
concept of nuclear deformation, hence of potential energy surfaces (PES) and 
fission barriers, is highly model-dependent: it is rooted in the mean-field 
approach to nuclear structure, and originates from the spontaneous symmetry 
breaking of rotational invariance in the intrinsic frame of the nucleus 
\cite{bender2003,ring2000}. However, the success of macroscopic-microscopic and 
self-consistent approaches in describing both qualitatively and quantitatively 
the main features of the fission process is evidence that such a concept is 
very useful in practice.


\subsubsection{Statistical Description of the Compound Nucleus}
\label{subsubsec-stat}

In a microscopic theory of fission based on nuclear DFT, it is assumed that 
the fission process is driven by a small set of collective degrees of freedom 
$\gras{q}$. It is further assumed that the potential energy surface of the 
compound nucleus in this collective space can be reliably described at the 
Hartree-Fock-Bogoliubov (HFB) approximation. This implies that the collective 
variables are defined as expectation values of specific operators, such as, 
e.g., multipole moments, on the HFB vacuum. The PES is then generated by 
performing a series of constrained HFB calculations. Such an approach is 
clearly an approximation, since the HFB vacuum is the lowest energy state for 
the set of constraints $\gras{q}$ while the compound nucleus is, by definition, 
in an excited state. Nonetheless, early calculations of fission fragment 
charge mass yields and total kinetic energy for low-energy neutron-induced 
fission obtained within this approximation give a reasonably good agreement 
with experimental data \cite{younes2012-a,goutte2005,berger1989}. In fact, 
a similar approximation is implicitly made in macroscopic-microscopic 
methods, with a similarly good reproduction of experimental data 
\cite{randrup2011,randrup2011-a}.

If the energy of the incident neutron $E_{n}$ increases (fast neutrons), the
excitation energy $E^{*}$ of the compound nucleus increases accordingly. For
$E_{n} \approx 14 $ MeV, $E^{*}$ can be typically of the order of 20 MeV or
more in actinides. In this regime, the nuclear level density is very large, 
of the order of $\rho(E^{*}) \approx 10^{12}$ MeV$^{-1}$ at $E^{*} \approx 20$ 
MeV and growing exponentially with $E^{*}$, see, e.g., Ref.~\cite{hilaire2012}. 
It thus becomes more and more unlikely that constrained HFB vacua can still 
provide a realistic description of the nuclear potential energy surface,
and more generally of the fission process. In addition, the extremely large
level density suggests that direct calculation of excited states could prove 
extremely challenging. Instead, we will seek to describe induced fission with 
finite-temperature density functional theory (FT-DFT).

In this work, we neglect particle evaporation or gamma emission. In other 
words, we do not consider second chance fission -- the fission of the nucleus 
after one neutron has been emitted, or third chance fission -- after two 
neutrons have been emitted. Therefore, the compound nucleus is viewed as a 
closed and isolated system. In statistical physics, such systems should be 
treated in the microcanonical ensemble \cite{reichl1988}. However, counting 
the number $\mathcal{N}(E^{*})$ of microstates of the system at any given 
experimental excitation energy $E^{*}$ would require one to have access to 
the full eigen-spectrum of the nucleus. In practice, this is impossible and 
the microcanonical treatment of the problem must be ruled out \cite{egido1988}.

In nuclear DFT, the nuclear wave-function takes the form of a HFB vacuum: 
it is not an eigenfunction of the nuclear Hamiltonian, nor of the particle 
number operator. This implies that the total energy and the number of 
particles in the system are only known on average: there can be fluctuations 
of both quantities, of either quantum or/and statistical origin 
\cite{egido1988}. This observation suggests to use the grand canonical 
ensemble to describe the nuclear system. The density operator $\hat{D}$ 
characterizing such an ensemble is obtained by maximizing Gibbs entropy 
under the constraints that the energy and particle numbers are constant on 
average. The resulting equation is equivalent to expressing the 
thermodynamical grand potential $\Omega$ at constant temperature $T$ and 
chemical potential $\lambda$ in terms of the grand partition function. The 
relevant thermodynamic potential is then the Helmholtz free energy $F$ 
\cite{reichl1988}. Note that, in this statistical setting, the temperature 
$T$ is, {\em stricto sensu}, only a Lagrange parameter used to maintain 
the energy constant on average.


\subsubsection{Neutron Incident Energy and Nuclear Temperature}
\label{subsubsec-En}

One of the difficulties in the FT-DFT description of fission is to 
interpret the temperature introduced in the theory, in particular in terms 
of the excitation energy of the compound nucleus. It was suggested in 
Refs.~\cite{goodman1991,moretto1972}  that the temperature $T$ be determined 
locally at every point in the collective space by assuming that all the 
excitation energy at deformation $\gras{q}$ is entirely of thermal nature. 
In practice, this scenario has only been applied in the 
macroscopic-microscopic approach to nuclear structure. Starting from the 
total energy at $T=0$ expressed as the sum of a macroscopic term, a shell 
correction and a pairing correction, one determines the local temperature 
$T(\gras{q})$ given the experimental excitation energy $E^{*}$; this 
temperature is then used to generate a new, temperature-dependent, PES 
where shell and pairing corrections are locally damped. This procedure 
has been used to describe superdeformed bands at high-spin and high 
excitation energy \cite{dudek1988}, hyperdeformation and the Jacobi 
shape transition \cite{schunck2007} and the dynamics of induced fission 
\cite{randrup2013}.

The feasibility of such an approach, however, is entirely contingent on 
the assumed decomposition of the energy into a temperature-independent 
part (the liquid drop energy) and a temperature-dependent microscopic 
correction, both of which depend on deformation. In FT-DFT, such a 
decomposition does not exist. All of the total energy is a function of 
the temperature: if one followed the recipe of selecting at point 
$\gras{q}$ in the collective space the DFT solution at $T=T(\gras{q})$ 
such that $E(\gras{q}) = E^{*}$, the total energy of the nucleus would 
become, by construction, constant across the collective space, and the 
very concept of a PES with barriers and valleys would be lost. 

\begin{figure}[!ht]
\center
\includegraphics[width=\linewidth]{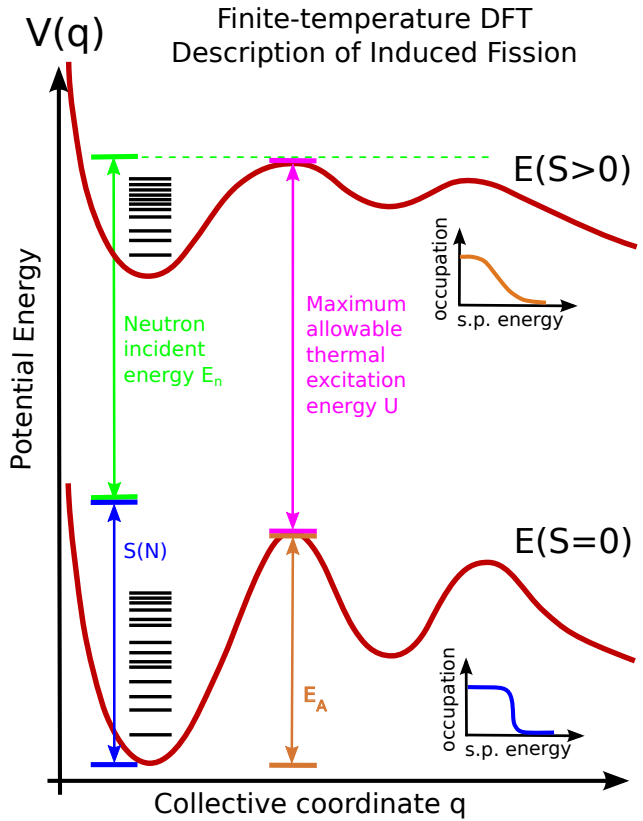}
\caption{(color online) Schematic illustration of how induced fission is
described within density functional theory at finite temperature. The
separation energy of the compound nucleus $(Z,N)$ is denoted $S(N)$.
}
\label{fig:schematic}
\end{figure}

In order to retain the view of fission as a large amplitude collective 
motion through a PES while simultaneously accounting for the effect of 
excitation energy via FT-DFT, we are thus bound to make the additional 
assumption that the temperature must be constant across the PES. More 
specifically, the potential energy surface can be defined either as 
the function $F(\gras{q}; T)$, where $F$ is the Helmholtz free energy, 
or by the function $E(\gras{q}; S)$, where $S$ is the entropy; see 
Sec. \ref{subsec-pathway} for additional details. In addition, since 
fission occurs for all neutron energies of interest, we will assume 
that the total excitation energy $E^{*}$ of the compound nucleus must 
be higher than the top of the barrier computed at $S>0$: this requirement 
gives us the maximum allowable {\it thermal} excitation energy $U$
available to the system. Figure~\ref{fig:schematic} illustrates how 
this works in practice: the thermal excitation energy is related to the 
entropy $S$ of the FT-HFB theory through 
$E_{n} + S(N) = U(S) + E_{\text{A}}$, where $E_{n}$ is the 
kinetic energy of the incident neutron, $S(N)$ is the one-neutron 
separation energy of the compound nucleus, and $E_{\text{A}}$ is the 
height of the first fission barrier at $S=0$. In calculations with 
energy functionals, the height of the first fission barrier may be 
larger than the separation energy, $E_{A} > S(N)$, which would 
contradict the experimental observation that the nucleus is fissile. 
For example, for $^{240}$Pu, we find $S(N) = 7.09$ MeV and 
$E_{\text{A}} = 7.65 $ MeV. In order to guarantee that thermal 
neutrons trigger fission, we thus have to introduce a small offset 
$\delta = E_{\text{A}} - S(N)$ such that 
$E_{n} + S(N) + \delta = U(S) + E_{\text{A}}$. In this particular 
case, $U(S) = E_{n}$. Note that this offset is a purely empirical 
correction needed to guarantee the fissile nature of the compound 
nucleus.

To finish this section, we note that the most rigorous way to combine a 
statistical description of the compound nucleus at high excitation energy 
with the conventional view of fission as a large amplitude collective 
motion would be to use the Liouville equation for the grand canonical 
density operator. Starting from some initial condition $\hat{D}_{0}$, 
the Liouville equation gives the time-evolution of $\hat{D}$. A 
collective, time-dependent equation of motion for the nucleus could then 
be obtained, at least in principle, by introducing the HFB approximation 
for the density operator and a small set of collective variables that would 
carry the time-dependence. Such a procedure was outlined in a recent paper, 
but numerous challenges remain to implement it in practice 
\cite{dietrich2010}.


\subsection{Finite Temperature HFB Theory}
\label{subsec-fthfb}

As recalled above, we use the finite-temperature HFB theory to describe the
compound nucleus at given excitation energy. The FT-HFB theory has a long
history in the literature
\cite{descloizeaux1968,blaizot1985,lee1979,goodman1981,tanabe1981,egido1993,egido2000,martin2003}. 
Here, we only recall the physical assumptions that are most relevant to this 
work. The compound nucleus is assumed to be in a state of thermal equilibrium 
at temperature $T$. In the grand canonical ensemble, the system is then
characterized by the statistical density operator $\hat{D}$,
\begin{equation}
\hat{D} = \frac{1}{Z} e^{-\beta(\hat{H} - \lambda\hat{N})},
\end{equation}
where $Z$ is the grand partition function, $\beta=1/kT$, $\hat{H}$ is the
Hamiltonian of the system, $\lambda$ the Fermi level and $\hat{N}$ the 
number operator \cite{blaizot1985,reichl1988}. In this work, the Hamiltonian 
is a two-body effective Hamiltonian with the Skyrme pseudopotential. The 
statistical average $\langle \hat{F} \rangle$ of an operator $\hat{F}$ is 
defined as
\begin{equation}
\langle\hat{F}\rangle = \text{Tr} \left[ \hat{D}\hat{F} \right],
\end{equation}
where the trace can be computed in any convenient basis of the Fock space,
i.e., it involves many-body states.

In the mean-field approximation of the density operator, the real Hamiltonian
$\hat{H}$ is replaced by a quadratic form $\hat{K}$ of the particle operators
\cite{blaizot1985,egido1993,egido2000,martin2003}. Given a generic basis 
$|i\rangle$ of the single-particle space, with $c_{i}$ and $c_{i}^{\dagger}$ 
the corresponding single-particle operators, this is expressed by
\begin{eqnarray}
\displaystyle\hat{K} & = & \sum_{ij} K_{ij}c_{i}^{\dagger}c_{j}\  (\text{HF}),
\label{eq:D_HF}
\medskip\\ 
\displaystyle\hat{K} & = &
\frac{1}{2}\sum_{ij} K_{ij}^{11} c_{i}^{\dagger}c_{j} +
\frac{1}{2}\sum_{ij} K_{ij}^{22} c_{i}c_{j}^{\dagger} \nonumber
\\ 
&& + \frac{1}{2}\sum_{ij} K_{ij}^{20} c_{i}^{\dagger}c_{j}^{\dagger}
+ \frac{1}{2}\sum_{ij} K_{ij}^{02} c_{i}c_{j}\ (\text{HFB}) 
\label{eq:D_HFB},
\end{eqnarray}
where HF refers to the Hartree-Fock approximation of the partition function, 
and HFB to the Hartree-Fock-Bogoliubov approximation. As a consequence of 
the Wick theorem for ensemble averages, there is a one-to-one correspondence 
between the one-body density matrix $\hat{\rho}$ (HF) or generalized density 
matrix $\hat{\mathcal{R}}$ (HFB) and the operator $\hat{K}$ 
\cite{blaizot1985,egido1993}. In particular: all statistical traces can be 
computed by taking the trace in the single-particle space \cite{blaizot1985},
\begin{equation}
\displaystyle
\langle\hat{F}\rangle 
=
\text{Tr} \left[ \hat{D}\hat{F} \right]
= \left\{
\begin{array}{l}
\text{tr} \left[ \hat{\mathcal{\rho}}\hat{F} \right],\  (\text{HF}),
\medskip\\ 
\frac{1}{2}\text{tr} \left[ \hat{\mathcal{R}}\hat{F} \right],\  (\text{HFB}). 
\end{array} \right.
\end{equation}

The forms (\ref{eq:D_HF})-(\ref{eq:D_HFB}) of the operator $\hat{K}$ 
defining the statistical density operator are generic. The matrix 
elements of $\hat{K}$ are thus taken as variational parameters and 
determined by requesting that the grand potential be minimum with 
respect to variations $\delta\hat{K}$. This leads to the identification 
$\hat{K} = \hat{h}$ (HF) and $\hat{K} = \hat{\mathcal{R}}$ (HFB), that 
is
\begin{eqnarray}
\displaystyle\hat{\rho} & = & \frac{1}{1 + \exp(\beta\hat{h})},\  (\text{HF}),
\label{eq:eq_HF}
\medskip\\ 
\displaystyle\hat{\mathcal{R}} & = & \frac{1}{1 + \exp(\beta\hat{\mathcal{H}})},\  (\text{HFB}),
\label{eq:eq_HFB}
\end{eqnarray}
where $\hat{h}$ is the usual HF Hamiltonian, $\hat{\mathcal{H}}$ 
the HFB Hamiltonian and $\beta = 1/kT$. These equations are the HF and HFB 
equations; see \cite{blaizot1985,egido1993} for the demonstration. Note 
that the variational principle does not require that either of these 
matrices be diagonalized. In practice, building the density matrix 
(generalized density) from the eigenvectors of $h$ ($\mathcal{R}$) just 
happens to be a very convenient way to guarantee that the functional
equations (\ref{eq:eq_HF})-(\ref{eq:eq_HFB}) are satisfied. 

In the basis where $\mathcal{H}$ is diagonal, one easily shows that the 
statistical occupation of a one quasi-particle state reads
\begin{equation}
\text{Tr} \left[ \hat{D}\beta_{\mu}^{\dagger}\beta_{\mu} \right]
= \frac{1}{1 + e^{\beta E_{\mu}}} \delta_{\mu\nu} = f_{\mu\nu}\delta_{\mu\nu},
\end{equation}
with $E_{\mu}$ the q.p. energy, i.e., the eigenvalue of $\mathcal{H}$. 
This result allows to show that the matrix of the one-body density matrix 
and pairing tensor in the s.p. basis are modified according to
\begin{eqnarray}
\rho_{ij}   & = & \text{Tr}\left[ \hat{D} c_{j}^{\dagger}c_{i} \right]
= \left( V^{*}(1-f)V^{T} \right)_{ij}
+ \left( UfU^{\dagger} \right)_{ij},
\label{eq:rho}\medskip\\
\kappa_{ij} & = & \text{Tr}\left[ \hat{D} c_{j}c_{i} \right]
= \left( V^{*}(1-f)U^{T} \right)_{ij}
+ \left( UfV^{\dagger} \right)_{ij},
\label{eq:kappa}
\end{eqnarray}
where the $U$ and $V$ are the matrices of the Bogoliubov transformation.

The finite-temperature extension of the HFB theory poses two difficulties.
First, we recall that, in the HFB theory at zero temperature, the component
$V_{\mu}$ of the q.p. $\mu$ is always localized for a system with negative
Fermi energy $\lambda < 0$ \cite{dobaczewski1984,dobaczewski1996}. The 
consequence is that both the mean-field, the pairing field, and the 
expectation value of any physical observable $\hat{O}$ are also localized 
(since $\rho= V^{*}V^{T}$ and $O = \text{tr} \hat{O}\hat{\rho}$). However, 
at finite-temperature, we note that, even though the pairing tensor remains 
always localized for $\lambda <0$, the density matrix does not. More 
specifically, all q.p. $\mu$ with $0 < E_{\mu} < -\lambda$ give a localized 
contribution to the mean-field and physical observables, while all q.p. with 
$E_{\mu} > -\lambda $ yield a coupling with the continuum through the 
$\left( UfU^{\dagger} \right)$ term of the density matrix, see 
Sec.~\ref{subsubsec-continuum} for more details. The existence of this 
coupling was already pointed out and quantified in the context of the 
Hartree-Fock theory at finite temperature \cite{bonche1985,bonche1984,brack1974}.

The second difficulty is that, in the statistical description of the system by
a grand canonical ensemble, only the average value of the energy and the
particle number (and any other constrained observables) are fixed. In addition
to the quantum fluctuations brought about by the fact that DFT wave-functions 
are not eigenstates of the Hamiltonian, thermal (or statistical) fluctuations 
are also present \cite{egido1988}. They increase with temperature and should 
decrease with the system size \cite{reichl1988}. From a statistical point of view, 
the FT-HFB theory only gives the most probable solution within the 
grand-canonical ensemble, the one that corresponds to the lowest free energy. 
Mean values and deviations around the mean values of any observable 
$\hat{\mathcal{O}}$ can be computed in the classical limit as in 
Ref.~\cite{martin2003}
\begin{equation}
\bar{\mathcal{O}} =
\frac{\displaystyle \int d^{N}\gras{q}\; \mathcal{O}(\gras{q})
e^{-\beta F(T,\gras{q})}}{\displaystyle \int d^{N}\gras{q}\;  e^{-\beta F(T,\gras{q})}}.
\end{equation}
Such integrals should in principle be computed across the whole collective
space defined by the variables $\gras{q} = (q_{1}, \dots, q_{N})$ and require
the knowledge of the volume element $d^{N}\gras{q}$. Other possibilities
involve functional integral methods \cite{levit1984}. In this work, we only
consider the most probable value for observables and disregard all statistical 
fluctuations.


\subsection{Skyrme EDF and Numerical and Numerical Implementation}
\label{subsec-numerics}

We briefly recall that we work with Skyrme energy densities, for which the
energy of the nucleus is a functional of the one-body density matrix. In this
paper, all calculations have been performed at the FT-HFB level with the SkM*
parametrization of the Skyrme pseudopotential \cite{bartel1982}. The pairing
functional originates from a density-dependent, mixed surface-volume pairing
force. In the calculations of the densities, all quasi-particles above a
cut-off energy $E_{\text{cut}} = 60$ MeV are dismissed. The pairing strength
for both the proton and neutron force were fitted locally on the 3-point
formula of the odd-even mass difference in $^{240}$Pu, see (I) for details.
Contrary to (I), the UNEDF family of functionals was not considered here, since
they require the Lipkin-Nogami prescription, which is not available yet at
finite temperature.

As in (I), the nuclear shape is characterized by a set
$\boldsymbol{q} = (q_{1},\dots,q_{N})$ of $N$ collective variables. In this
work, we consider the expectation value $q_{\lambda\mu}$ of the multipole
moment operators $\hat{Q}_{\lambda\mu}$ on the HFB vacuum for the: axial
quadrupole $(\lambda,\mu)=(2,0)$; triaxial quadrupole $(\lambda,\mu)=(2,2)$;
axial octupole $(\lambda,\mu)=(3,0)$ and axial hexadecapole
$(\lambda,\mu)=(4,0)$. We also employ the expectation value of the neck
operator $\hat{Q}_{N}$ with the range $a_{N} = 1.0$ fm. The finite-temperature
extension of the Wick theorem guarantees that the expectation value of these
(one-body) operators at $T>0$ take the same form as at $T=0$, only with the
density matrix computed as in (\ref{eq:rho}). Constrained HFB solutions are
obtained by using a variant of the linear constraint method where the Lagrange
parameter is updated based on the cranking approximation of the random phase
approximation (RPA) matrix \cite{decharge1980,younes2009,schunck2011}. This 
method has been extended to handle non-zero temperatures.
All calculations were performed with the DFT solvers HFODD \cite{schunck2011} and
HFBTHO \cite{stoitsov2013}. In both codes, the HFB eigenfunctions are expanded on
a one-center harmonic oscillator  (HO) basis. In all calculations reported here,
this expansion was based on the lowest $N_{\text{states}} = 1100$ states of the
deformed HO basis. The largest oscillator shell entering the expansion was
$N_{\text{max}} = 31$. The deformation $\beta_{2}$ and the oscillator frequency
$\omega_{0}$ of the HO were parametrized empirically as a function of the
requested  expectation value $q_{20}$ of the quadrupole moment $\hat{Q}_{20}$
according to
\begin{equation}
\omega_{0} =
\left\{ \begin{array}{l}
0.1 \times q_{20}e^{-0.02 q_{20}} + 6.5 \text{MeV}\ 
\text{if}\ |q_{20}| \leq 30 \text{b} \\
8.14 \text{MeV} \ \text{if}\ |q_{20}| > 30 \text{b}
\end{array}
\right.
\label{eq:omega}
\end{equation}
and
\begin{equation}
\beta = 0.05\sqrt{q_{20}}
\label{eq:beta}
\end{equation}
We refer the reader to (I) and Ref.~\cite{schunck2013-a} for further details 
on the convergence properties of the basis.


\section{Evolution of Deformation Properties at Finite Temperature}
\label{sec-pathway}

In this section, we illustrate the modifications of the collective potential
energy surfaces discussed in (I) induced by the finite temperature. In
particular, we give an accurate estimate of the evolution of fission barrier
heights as a function of the excitation energy of the compound nucleus formed in
the reaction $^{239}$Pu(n,f).


\subsection{Fission Pathway of Least Energy}
\label{subsec-pathway}

As recalled in Sec.~\ref{subsec-fthfb} the FT-HFB theory is built on the
grand-canonical description of the nucleus as a system in thermal equilibrium
maintained at constant temperature $T$. Since particle number is constant on
average across the whole collective space, the thermodynamical potential
relevant to study deformation effects is the Helmholtz free energy
$F = E - TS$, which is computed at constant volume $V$ and temperature $T$. The
potential energy surface is characterized by the ensemble of points
$F(\gras{q})$, and variations of free energy between two points $\gras{q}_{1}$
and $\gras{q}_{2}$ are computed through
$\delta F|_{T} = F(\gras{q}_{1},T) - F(\gras{q}_{2},T)$.

\begin{figure}[!ht]
\center
\includegraphics[width=\linewidth]{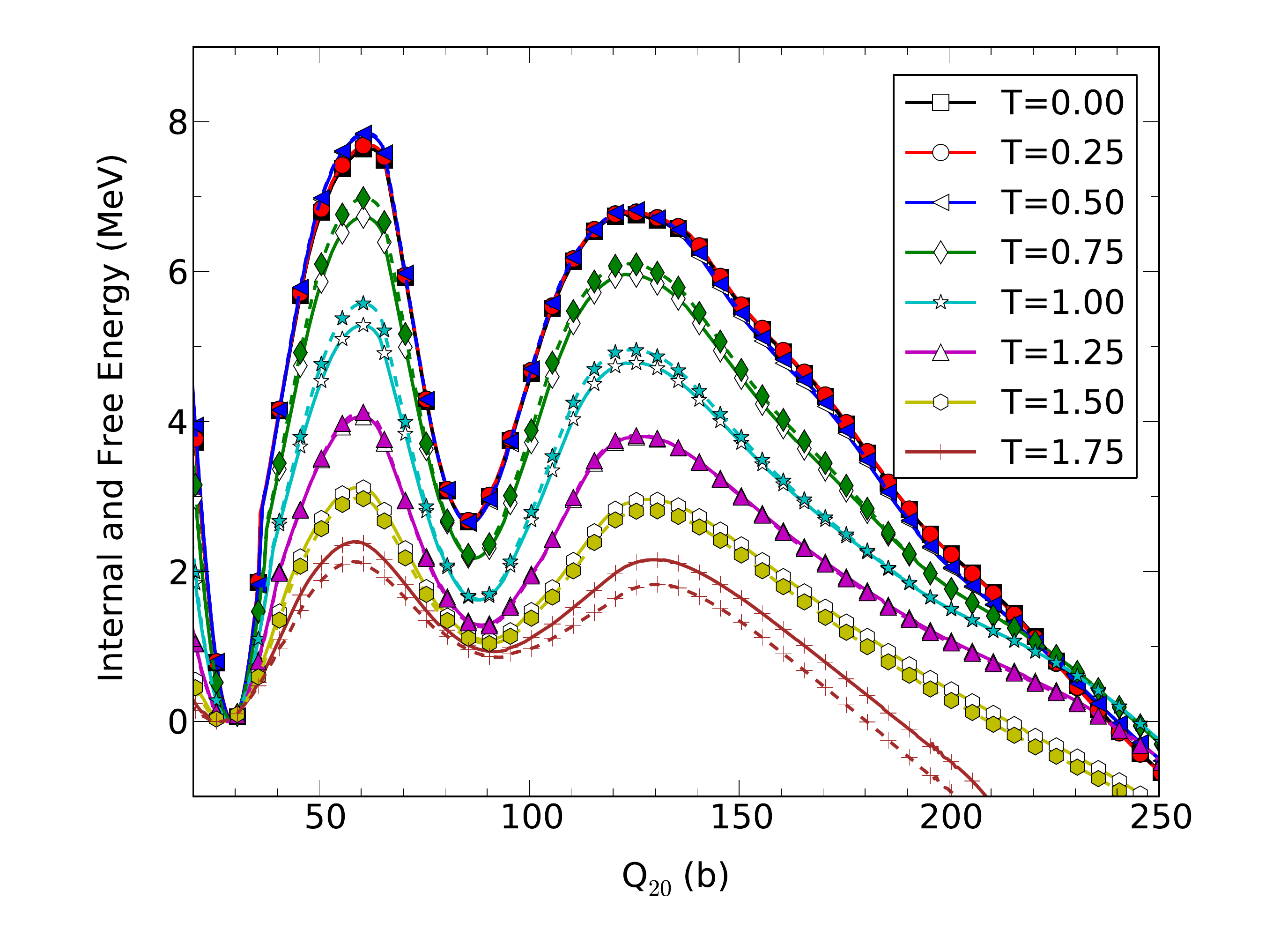}
\caption{(color online) Plain lines with open symbols: Free energy along the
least-energy fission pathway in $^{240}$Pu for finite temperatures
$T=0.00, \dots,1.75$ MeV. Dashed lines with plain symbols: Corresponding
internal energy $E$ at constant entropy $S$. All curves are normalized to their
ground-state value. Temperature units are in MeV.
}
\label{fig:PES_1D_energyT}
\end{figure}

We show in Fig.~\ref{fig:PES_1D_energyT} the free energy of the compound
nucleus $^{240}$Pu along the least-energy fission pathway for temperatures
ranging between 0 and 1.75 MeV by steps of 250 keV. Based on the discussion 
of Sec.~\ref{subsec-thermo}, this corresponds to maximal excitation energies 
of about $E^{*} \approx 80$ MeV for the compound nucleus. Recall that the
height of the first fission barrier is $E_{\text{A}} = 7.65$ MeV in $^{240}$Pu
for the SkM* functional; this corresponds to maximum neutron kinetic energies
of about $E_{n} \approx 73$ MeV. The least-energy fission pathway is found
according to the procedure presented in (I): while the value of the axial
quadrupole moment is constrained, the triaxial, octupole and hexadecapole
moments are unconstrained, so that triaxiality and mass asymmetry effects are
taken into account.

It has been argued in the literature that an isentropic description of fission
should be preferred over the isothermal description \cite{pei2009,diebel1981}. In
this representation, the thermodynamical potential is the internal energy $E$,
which is computed at constant volume $V$ and entropy $S$. The potential energy
surface is now the ensemble of points $E(\gras{q})$, and variations of energy
are computed through $\delta E|_{S} = E(\gras{q}_{1};S) - E(\gras{q}_{2};S)$.
The Maxwell relations of thermodynamics state that the variations of the free
energy $\delta F|_{T}$ over some extensive state variable $X$ (at constant
temperature) are equal to the variations of the internal energy $\delta E|_{S}$
(at constant entropy) \cite{reichl1988}.

Figure~\ref{fig:PES_1D_energyT} also shows the internal energy $E(\gras{q};S)$
at constant entropy deduced from the free energy curves: At each value
$\gras{q}$ of the constrained collective variable (here, the axial quadrupole
moment $\hat{Q}_{20}$), the quantities $E(\gras{q};T)$ and $S(\gras{q};T)$ are
used to reconstruct the relation $E(\gras{q};S)$ by regression. For each
temperature $T$, the curves $E(\gras{q};S)$ are then generated by fixing the
entropy at its value at the top of the second barrier for that temperature $T$.
Note that we could choose the entropy at other deformations: when properly
normalized, the Maxwell relations guarantee that all these choices should be
strictly equivalent, within the numerical accuracy of the regression.
Figure~\ref{fig:PES_1D_energyT} indicates that this accuracy is of the order of
500 keV at worst. This confirms earlier calculations \cite{pei2009,schunck2013}.

The isentropic representation of the fission process is often thought of as
more physically justified than the isothermal one, as it has its origin in the
separation of scales between the slow collective motion and fast intrinsic
excitations of the nucleus \cite{baranger1978,dietrich2010}. This separation justifies
the thermodynamical assumption of adiabaticity \cite{landau1980}: going from point
$\gras{q}$ to point $\gras{q'}$ in the collective space can be accomplished via
a quasi-static, reversible transformation that conserves entropy. By contrast,
it is sometimes argued that the absence of a heat bath to maintain the
temperature constant invalidates the isothermal representation \cite{diebel1981}.

Such a statement, however, comes from a misconception about the nature of the
heat bath. Indeed, another way to interpret the separation of scales between
collective and intrinsic motion is to write the energy density of the
fissioning nucleus as
\begin{equation}
\mathcal{H} = \mathcal{H}_{\text{coll}}(\gras{q}) + \mathcal{H}_{\text{int}},
\label{eq:collective}
\end{equation}
with the collective part depending only on the collective coordinates
$\gras{q}$ while the intrinsic part depends on all intrinsic degrees of
freedom. In a DFT picture, for example, we would take
$\mathcal{H}_{\text{int}} = \mathcal{H}_{\text{int}}[\rho,\kappa]$. The number
of intrinsic degrees of freedom is given by the value of $\rho$ and $\kappa$ at
every point in space, spin and isospin space: it is considerably higher than
the number of collective variables. In addition, in the limit of no
dissipation, the couplings between the two types of motion can be neglected
\cite{dietrich2010}. The decomposition (\ref{eq:collective}), together with the
different relaxation scales, shows that the role of the heat bath is in fact
played by the intrinsic Hamiltonian. In the theory of quantum dissipation, 
the latter is often modeled by a collection of harmonic oscillators
\cite{dittrich1998,razavy2005}. Passing from point $\gras{q}$ to point $\gras{q'}$ can
thus also be accomplished through an isothermal process, during which heat will
be exchanged between the collective wave-packet and the intrinsic excitations,
according to $\delta Q = TdS$. In our opinion, the two representations, which
are mathematically equivalent thanks to the Maxwell equations, are also
physically equivalent since they only rely on the hypothesis of the separation
of degrees of freedom into slow collective and fast intrinsic motion.

To conclude this section, we note that, in order for the Maxwell relations to
be valid, the respective thermodynamical potentials $F(\gras{q})|_{T}$ and
$E(\gras{q})|_{S}$ must be differentiable at point $\gras{q}$. As discussed
in Sec.~\ref{subsec-fragments}, this may not be true near scission, at least 
in the 4-d collective space explored in Fig.~\ref{fig:PES_1D_energyT}.


\subsection{Dependence of Fission Barriers on Excitation Energy}
\label{subsec-barriers}

\begin{figure}[!ht]
\center
\includegraphics[width=\linewidth]{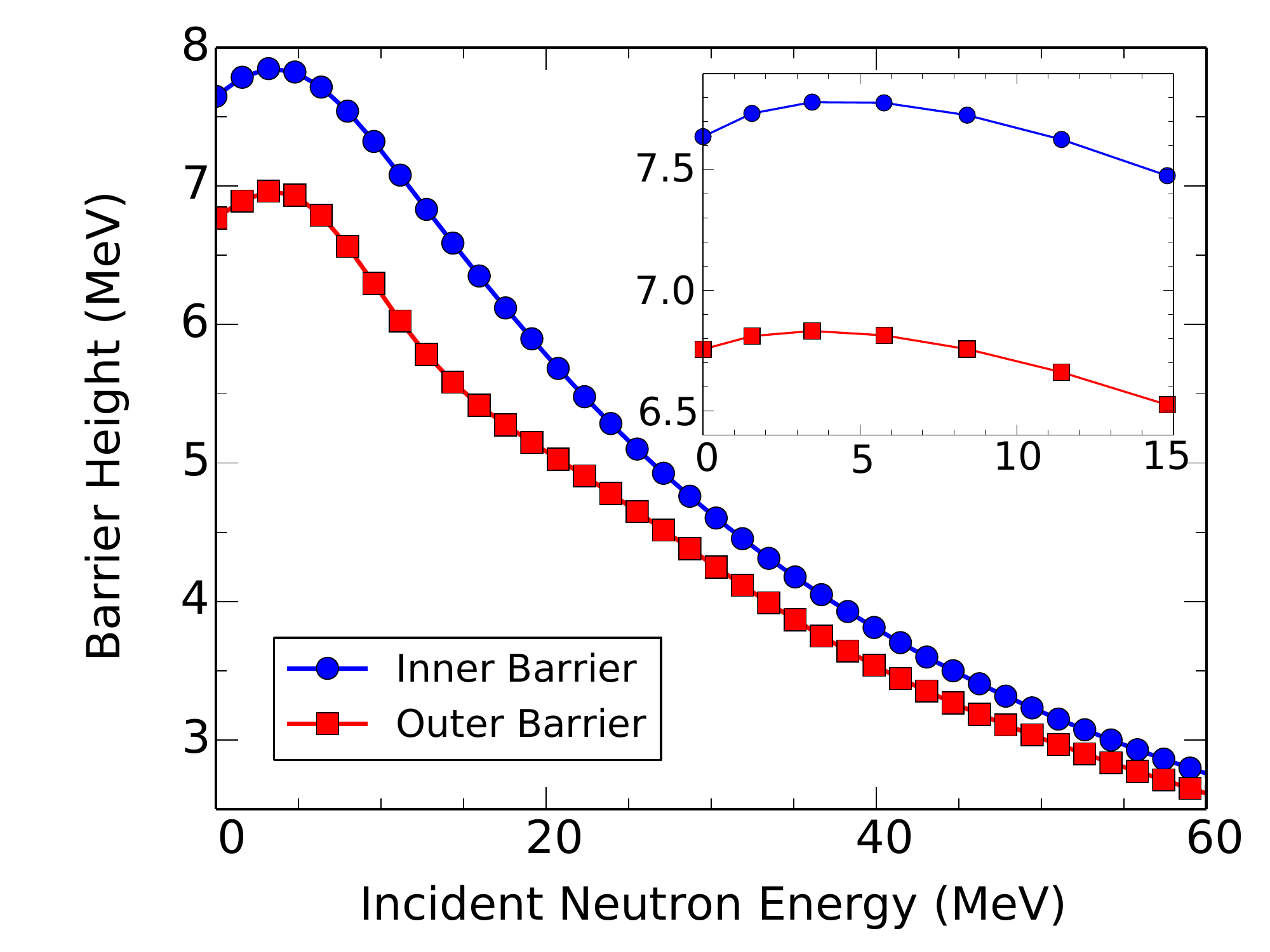}
\caption{(color online) Evolution of the inner and outer barrier heights in
$^{240}$Pu as a function of the energy of the incident neutron. The inset 
represents a close-up in the 0 -- 15 MeV region.
}
\label{fig:PES_1D_barriersT}
\end{figure}

Fission barrier heights (both inner and outer barriers) are particularly
important quantities in fission models, as they are often used as input to
reaction codes. In Fig.~\ref{fig:PES_1D_barriersT}, we show the variation of
the inner and outer fission barrier heights in $^{240}$Pu as a function of 
the incident neutron energy $E_{n}$. As outlined in Sec.\ref{subsec-thermo}, 
we compute the fission barrier at incident neutron energy $E_{n}$ for the 
entropy $S$ such that $E_{n} = U(S)$. The maximum allowable thermal excitation 
energy is deduced from the $E(\gras{q})|_{S}$ curves, which are obtained by 
spline interpolation over the $F(\gras{q})|_{T}$ calculations. Note that at 
any point $\gras{q}$, the error on the total energy at given entropy due to 
the interpolation is smaller than 50 keV.

In the literature, fission barriers at finite temperature were computed 
within the macroscopic-microscopic approach 
\cite{ignatyuk1980,sauer1976,mosel1974,hasse1973}, 
the semi-classical Thomas-Fermi framework
\cite{garcias1990,garcias1989,guet1988,dalili1985,nemeth1985,diebel1981}, 
and the self-consistent HF theory \cite{bartel1985,nemeth1985}. There are 
also a few applications of the finite-temperature HFB theory with both 
zero-range Skyrme functionals and finite-range Gogny forces 
\cite{pei2009,sheikh2009,martin2009}. All these studies point to the 
disappearance of the barriers with the excitation energy of the compound 
nucleus, or equivalently, the nuclear temperature. Our results confirm 
this overall trend.


However, we emphasize here that this phenomenon occurs at temperatures 
that are relatively high as far as applications of neutron-induced 
fission are concerned. In the regime $0\leq E_{n} < 5-6$ MeV, the 
somewhat unexpected effect of nuclear temperature is to slightly 
{\it increase} fission barriers. In the inset of figure  
\ref{fig:PES_1D_barriersT}, we show a close-up of the fission barrier 
heights in the region $0\leq E_{n} < 15$ MeV. There is a very clear 
upward trend at low neutron energies. Although the increase of the 
fission barriers does not exceed 200 keV, the effect may be significant 
enough to affect fission fragment distributions.

\begin{figure}[!ht]
\center
\includegraphics[width=\linewidth]{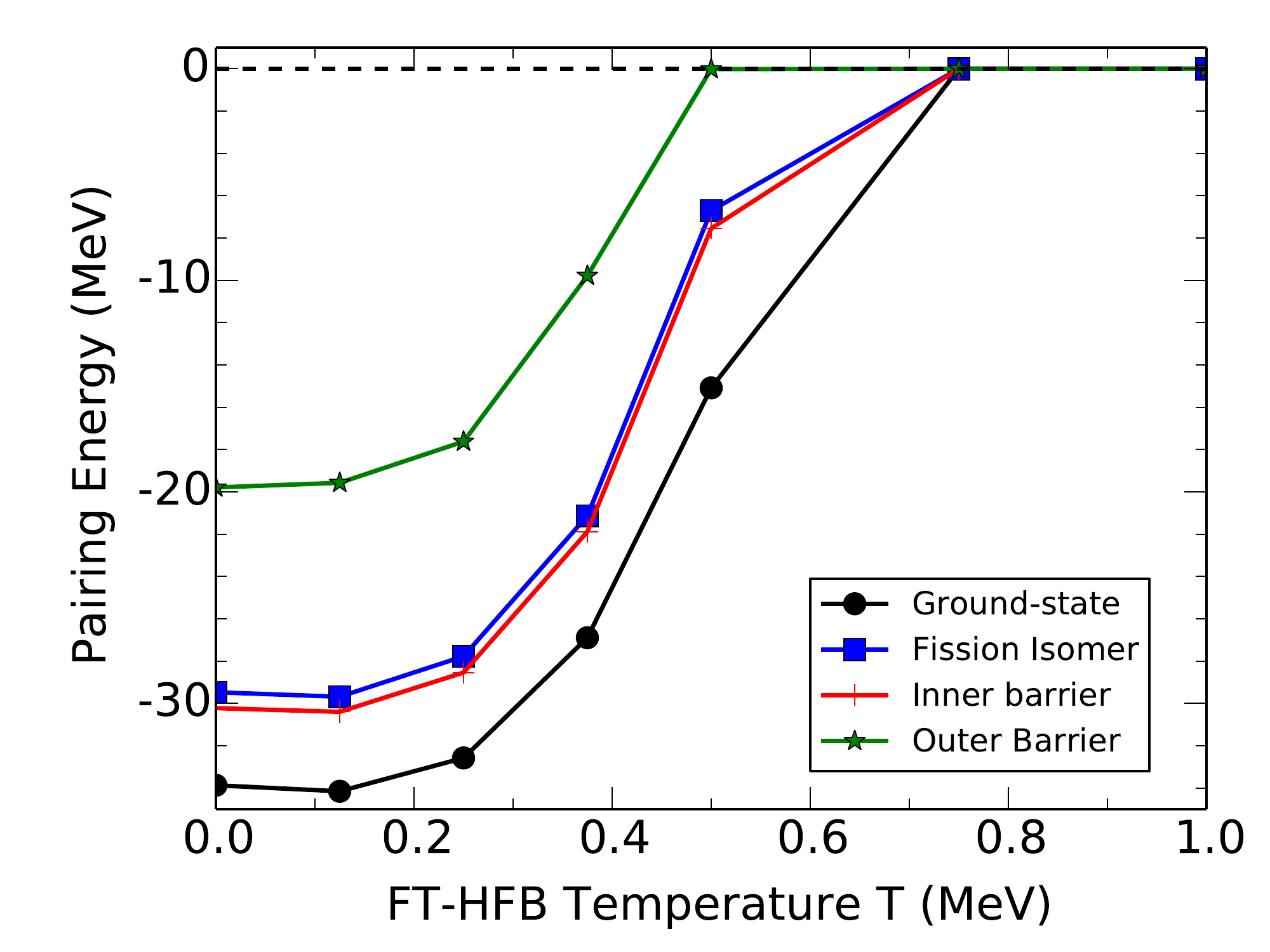}
\caption{(color online) Evolution of pairing energy in the ground-state, 
fission isomer, at the top of the first barrier and at the top of the 
second barrier in $^{240}$Pu as a function of the FT-HFB nuclear 
temperature.
}
\label{fig:PES_1D_barriersT_Epair}
\end{figure}

The reason for the counter-intuitive behavior of the barriers may be 
attributed to the different damping speeds of pairing correlations and 
shell effects with temperature. In figure \ref{fig:PES_1D_barriersT_Epair}, 
we show the pairing energy as a function of the FT-HFB temperature for 
the ground-state, fission isomer, top of the first barrier and top of 
the second barrier in $^{240}$Pu. We find that pairing correlations 
have vanished completely beyond $T=0.75$ MeV, which corresponds to a 
neutron energy of approximately 12 MeV. However, shell effects are 
still substantial at this temperature \cite{brack1974}. Our 
interpretation is that the fast damping of pairing correlations 
attenuates, delays, or even partially reverts the impact of the damping 
of shell effects on the deformation energy as a function of temperature. 
Indeed, one of the side-effects of pairing correlations is to reduce 
deformation energy \cite{brack1972,moretto1972}, i.e., the energy 
difference $E_{\text{def}}(\gras{q}) = E(\gras{q}) - E(\gras{q}=0)$. 
If pairing correlations rapidly decrease as a function of $T$, the 
absolute value of the deformation energy may slightly increase as a 
result. Of course, this qualitative interpretation should be validated 
by rigorous macroscopic-microscopic calculations.


\section{Fission Fragment Properties at Finite Temperature}
\label{sec-scission}

In (I), we discussed the fission fragment properties of $^{239}$Pu(n,f) at
$T=0$ using the Joint Contour Net (JCN) to define a scission region in terms of
topological changes of the density -- based on the assumption that the
variations of the density in the pre-fragments must be commensurate with those
of the density in the compound nucleus. Within this region of scission
configurations, we then apply a quantum localization method to disentangle the
pre-fragments in order to approach the asymptotic conditions of two fully
independent fragments. In this section, we extend this study to finite
temperature.


\subsection{Definition of a Scission Region}
\label{subsec-region}

The fission pathway of lowest free energy across the 4-dimensional collective
space shown in Fig.~\ref{fig:PES_1D_energyT} was extended up to the scission
region for each temperature. We find that the value $q_{20}^{(\text{disc})}$ 
of the axial quadrupole moment where the first discontinuity in the $F(q_{20})$
curve appears changes with temperature. Table~\ref{tab:scission} lists these
values as a function of the temperature for the SkM* functional. Note that 
there is a numerical uncertainty of about 2--3 b for the values of $q_{20}$ 
reported in the table, since calculations converge very slowly near scission. 
The scission region covers a relatively broad range of $\hat{Q}_{20}$ values 
of approximately 25 b. Note the original increase of the quadrupole moment at 
low temperatures: this is caused by the quenching of pairing correlations, 
which was shown to shift the discontinuity at larger values of $\hat{Q}_{20}$ 
in (I). Since the discontinuity does not occur at the same $q_{20}$ for all
temperatures, the equivalence between the $E(q_{20})|_{S}$ and  $F(q_{20})|_{T}$ 
representations of the fission pathway does not hold in the scission region 
since neither the internal energy nor the free energy are continuous functions 
over the entire range of quadrupole moments involved.

\begin{table}[!ht]
\begin{center}
\caption{Position of the first discontinuity of the $F(q_{20})$ curve
along the most probably fission pathway as a function of temperature.}
\begin{ruledtabular}
\begin{tabular}{cccc}
T (MeV) & $q_{20}^{(\text{disc})} (b)$ & T (MeV) & $q_{20}^{(\text{disc})} (b)$   \\
\hline
0.00 & 345.0 & 1.00 & 332.5 \\
0.25 & 357.0 & 1.25 & 331.0 \\
0.50 & 339.5 & 1.50 & 333.0\\ 
0.75 & 332.0 & 1.75 & 333.0\\
\end{tabular}
\label{tab:scission}
\end{ruledtabular}
\end{center}
\end{table}

\begin{figure}[!ht]
\center
\includegraphics[width=0.9\linewidth,angle=+90]{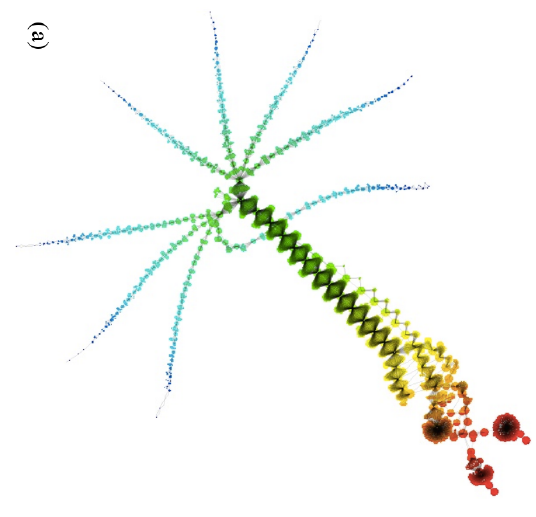}
\includegraphics[width=0.9\linewidth,angle=+90]{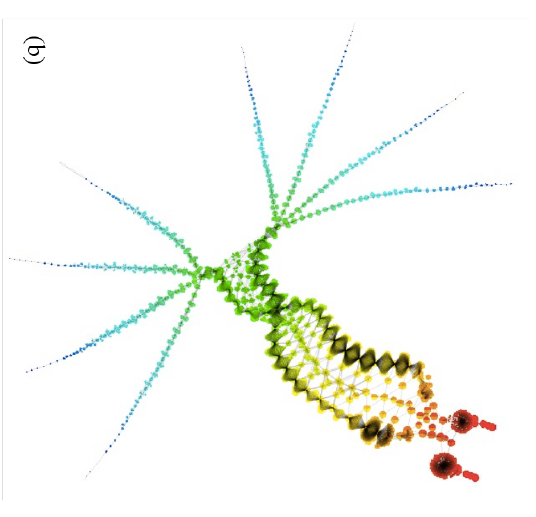}
\caption{(color online) Joint Contour Net graph of the densities at
$q_{N} = 0.2$ at temperature $T=0.00$ MeV (a) and $T = 1.50$ MeV (b).
}
\label{fig:JCN}
\end{figure}

Following the approach at zero-temperature outlined in (I), we introduce an
additional constraint on the number of particles in the neck, $\hat{Q}_{N}$, 
to explore scission configurations. At each temperature $T$, the expectation 
value of $\hat{Q}_{N}$ is varied in the range $q_{N} \in [0.1,4.5]$, while 
the quadrupole moment is fixed at the values listed in Table \ref{tab:scission}.
The JCN analysis is then applied at each temperature to the set of neutron 
and proton densities along these trajectories to identify putative scission
configurations. As illustration, figure \ref{fig:JCN} shows the JCN at 
$q_{N} = 0.2$ at both $T=0.0$ MeV (top) and $T=1.5$ MeV (bottom).

We recall that the JCN algorithm provides a computational tool for extracting
the topological features of a multifield dataset, which includes connectivity
between regions of different behaviors. It was first introduced in the context
of nuclear structure in Ref.~\cite{duke2012} and applied in (I) to the specific
problem of defining scission configurations along a continuous fission pathway
for neutron-induced fission. The JCN analysis involves generalizing
one-dimensional scalar analysis to capture simultaneous variation in multiple
output functions of the type
$(f_{1}, \dots, f_{n}): \mathbb{R}^{3} \rightarrow \mathbb{R}^{n}$. In (I), 
we concluded that the JCN could be a useful tool to define plausible scission
configurations. In particular, the appearance of a branching structure in the
JCN, which characterizes the existence of two distinct regions in space, was
interpreted as the precursor to scission; the subsequent development of
``starbursts'' in each branch was associated with the completion of scission,
as these startbursts indicate that the variations of the density in each
fragment is commensurate to those of the density in the whole nucleus (hence,
suggesting two well-defined fragments). We also observed that this
identification is independent of the numerical parameters used in the JCN.

\begin{table}[!ht]
\begin{center}
\caption{Interval $I_{q} = [ q^{\text{(min)}}, q^{\text{(max)}}]$ of scission
configurations as obtained from the JCN analysis of $^{240}$Pu as a function of
the temperature $T$.}
\begin{ruledtabular}
\begin{tabular}{cccccc}
T (MeV) & $q^{\text{(min)}}$ & $q^{\text{(max)}}$ & 
T (MeV) & $q^{\text{(min)}}$ & $q^{\text{(max)}}$\\
\hline
0.00 & 0.2  & 2.6 & 1.00 & 0.3  & 3.1 \\
0.25 & 0.2  & 2.5 & 1.25 & 0.5  & 3.1 \\
0.50 & 0.2  & 2.8 & 1.50 & 0.5  & 3.1 \\ 
0.75 & 0.2  & 3.0 & 1.75 & 0.7  & 3.1 \\
\end{tabular}
\label{tab:JCN}
\end{ruledtabular}
\end{center}
\end{table}

The results of the JCN analysis for the least energy fission pathway of
$^{240}$Pu at finite temperature are summarized in Table~\ref{tab:JCN}. As 
in (I), we define an interval $I_{q}=[ q^{\text{(min)}}, q^{\text{(max)}}]$
in the collective space, with $q^{\text{(min)}}$ the value of $\hat{Q}_{N}$ 
where scission has completed (=the actual scission point), and 
$q^{(\text{max)}}$ the value corresponding to the precursor to scission. We 
note that the precursor value is relatively stable, especially at high 
temperatures, while the position of the scission point, which is constant 
up to $T =1.0$ MeV, moves to thicker necks beyond $T > 0.75$ MeV. We will 
return to this result in Sec.~\ref{subsec-fragments}.

The JCN also picked up an interesting ``zippering effect'' of the datasets (the
proton and neutron densities) at large temperatures and low $q_{N}$ values.
This effect is illustrated in Fig.~\ref{fig:JCN}, which shows the JCN at
$q_{N} = 0.2$ at both $T=0.0$ MeV (top) and $T=1.5$ MeV (bottom). In both
cases, the fragments are clearly formed as evidenced by the two distinct
branches in the upper right side of each figure. In addition, we notice at
$T=1.5$ MeV a complex pattern connecting the two fragments, which look similar
to a zipper. We have found that this pattern becomes more noticeable for
$T\geq 1.25$ MeV. Since the zippering connects the two pre-fragments, it should
be indicative of a spatial connection between these two distinct regions of
space; in addition, the effect manifests itself only at temperatures where the
coupling to the continuum becomes sizable, see Sec.~\ref{subsubsec-continuum}.
Therefore, we suggest that the zippering effect of the JCN is the
representation of a spatial delocalization of quasiparticles (mostly neutrons) 
at large temperatures.


\subsection{Quasi-Particle Occupations}
\label{subsec-occupations}

The generalization at $T>0$ of the procedure to identify a left and a right
fragment, their observables and their interaction energy presented in (I) is
straightforward. Using the definition (\ref{eq:rho}) for the one-body density
matrix at finite temperature $T>0$, we find that the coordinate space 
representation $\rho_{\mu}(\gras{r}\sigma,\gras{r}'\sigma')$ of the density  
of a single quasi-particle $\mu$ reads
\begin{multline}
\rho_{\mu}(\gras{r}\sigma,\gras{r}'\sigma') 
= 
\sum_{ij}
\left[ 
V_{i\mu}^{*}(1-f_{\mu})V_{j\mu} + U_{i\mu}f_{\mu}U_{j\mu}^{*} 
\right] \\
\times \phi_{i}(\gras{r}\sigma)\phi_{j}^{*}(\gras{r}'\sigma'),
\label{eq:occupation}
\end{multline}
with $\phi_{i}(\gras{r}\sigma)$ the single-particle basis functions. With 
this definition, the spatial occupation $N_{\mu}$ of the q.p. $\mu$ and the 
total number of particles $N$ are formally the same as at $T=0$, that is,
\begin{equation}
N_{\mu} = \sum_{\sigma} \int d^{3}\gras{r}\; \rho_{\mu}(\gras{r}\sigma,\gras{r}\sigma),
\end{equation}
and
\begin{equation}
N_{\mu} = \sum_{i} \left[ 
V^{*}_{i\mu} (1-f_{\mu}) V_{i\mu} + U_{i\mu}f_{\mu} U_{i\mu}^{*}.
\right ]
\label{eq:occ}
\end{equation}
As in (I), we introduce the quantity
\begin{equation}
d_{ij}(z) 
= 
\sum_{\sigma}
\int_{-\infty}^{+\infty} dx\int_{-\infty}^{+\infty} dy \int_{-\infty}^{z}dz\;
\phi_{i}(\gras{r}\sigma)\phi_{j}^{*}(\gras{r}\sigma).
\end{equation}
Still assuming that the neck between the two fragments is located on the
$z$-axis of the intrinsic reference frame, and thus has the coordinates
$\gras{r}_{\text{neck}} = (0,0, z_{N})$, we can define the occupation of 
the q.p. in the fragment (1) as
\begin{equation}
N_{1,\mu} 
= 
\sum_{ij}
\left[ 
V_{i\mu}^{*}(1-f_{\mu})V_{j\mu} + U_{i\mu}f_{\mu}U_{j\mu}^{*} 
\right] d_{ij}(z_{N}),
\label{eq:occupation1}
\end{equation}
As at $T=0$, the occupation of the q.p. in the fragment (2) is simply
$N_{2,\mu} = N_{\mu} - N_{1,\mu}$. We then assign the q.p. $\mu$ to 
fragment (1) if $N_{1,\mu} \geq 0.5 N_{\mu}$, and to fragment (2) if
$N_{1,\mu} <0.5 N_{\mu}$. This gives us two sets of quasiparticles. For 
each of them, we can define the corresponding pseudodensities and pseudo 
pairing tensor. For fragment (f) we find, in coordinate$\otimes$spin 
space,
\begin{eqnarray}
\rho^{(\text{f})}(\gras{r}\sigma,\gras{r}'\sigma') 
& = & 
\sum_{\mu\in (f)}\sum_{ij}
\left[ 
V_{i\mu}^{*}(1-f_{\mu})V_{j\mu} 
\right. \nonumber\\
& & \left.
+ U_{i\mu}f_{\mu}U_{j\mu}^{*} 
\right] \phi_{i}(\gras{r}\sigma)\phi_{j}^{*}(\gras{r}'\sigma'), 
\label{eq:rho_f_r}\medskip \\
\kappa^{(\text{f})}(\gras{r}\sigma,\gras{r}'\sigma') 
& = & 
\sum_{\mu\in (f)}\sum_{ij}
\left[ 
V_{i\mu}^{*}(1-f_{\mu})U_{j\mu} 
\right. \nonumber\\
& & \left. 
+ U_{i\mu}f_{\mu}V_{j\mu}^{*} \right]
\phi_{i}(\gras{r}\sigma)\phi_{j}^{*}(\gras{r}'\sigma').
\label{eq:kappa_f_r}
\end{eqnarray}
These are the equivalent at $T>0$ of Eqs.(12)-(13) in (I). We can build the 
analog of the kinetic energy density and the spin current tensor from these 
pseudodensities. The Coulomb and nuclear interaction energy between the two 
fragments thus takes the same form as at $T=0$, only the definition of the 
various pseudodensities is modified according to Eq.(\ref{eq:rho}) and 
Eq.(\ref{eq:rho_f_r}).


\subsection{Quantum Localization at Finite Temperature}
\label{subsec-localization}

As recalled in (I), at $T=0$ the localization method of Ref.~\cite{younes2011} 
is based on the idea that any unitary transformation of the q.p. operators
$(\beta^{\dagger},\beta)$ leaves the generalized density matrix, hence all
global observables such as the total energy, radii, etc., invariant. In this
section, we generalize this result at $T>0$ and discuss how it impacts the 
practical implementation of the method.


\subsubsection{Unitary Transformation of Quasiparticles}
\label{subsubsec-transformation}

As in (I), we consider the following unitary transformation $\hat{T}$ of the 
eigenvectors of the HFB matrix,
\begin{eqnarray}
A_{\alpha} & \equiv & \hat{T}U_{\mu} = \sum_{\mu} T_{\alpha\mu} U_{\mu},
\label{eq:unitary_00} \\
B_{\alpha} & \equiv & \hat{T}V_{\mu} = \sum_{\mu} T_{\alpha\mu} V_{\mu}
\label{eq:unitary_01}
\end{eqnarray}
where the quantities $A_{\alpha}$, $B_{\alpha}$, $U_{\mu}$ and $V_{\mu}$ are in
fact vectors with $N$ components $A_{n\alpha}$ in the original s.p. basis, so
that, in matrix form,
\begin{equation}
A = UT^{T}, \ \ B = VT^{T}.
\end{equation}
It is straightforward to notice that the matrix $\mathcal{W}'$ defined as
\begin{equation}
\mathcal{W}' =
\left( \begin{array}{cc}
A & B^{*} \\
B & A^{*}
\end{array}\right)
=
\left( \begin{array}{cc}
U & V^{*} \\
V & U^{*}
\end{array}\right)
\mathcal{T}^{\dagger}
=
\mathcal{W}\mathcal{T}^{\dagger},
\label{eq:unitary_implemented}
\end{equation}
with 
\begin{equation}
\mathcal{T} = 
\left( \begin{array}{cc}
T^{*} & 0 \\
0 & T
\end{array}\right), \ \ \mathcal{T}\mathcal{T}^{\dagger} = \mathcal{T}^{\dagger}\mathcal{T} = 1,
\end{equation}
verifies $\mathcal{W}'\mathcal{W}'^{\dagger}=\mathcal{W}'^{\dagger}\mathcal{W}'=1$.
The matrices $A$ and $B$ thus define new sets of q.p. operators
$(\eta^{\dagger}, \eta)$ such that
\begin{equation}
\eta_{\alpha}           = \sum_{\mu} T^{*}_{\alpha\mu}\beta_{\mu}, \ \ \
\eta_{\alpha}^{\dagger} = \sum_{\mu} T_{\alpha\mu}\beta_{\mu}^{\dagger}.
\label{eq:unitary1}
\end{equation}
Therefore, a unitary transformation of the Bogoliubov matrices of the type
(\ref{eq:unitary_00})-(\ref{eq:unitary_01}) correspond to a transformation of 
the q.p. creation (annihilation) operators into linear combination of 
themselves without mixing creation and annihilation operators \cite{blaizot1985}.

Using the Baker-Hausdorff Campbell formula, it is not very difficult to show
that, for the form (\ref{eq:unitary1}) of the unitary transformation
there exists in Fock space a general transformation rule for the q.p. operators
\begin{equation}
\eta_{\mu}           = e^{i\hat{R}} \beta_{\mu} e^{-i\hat{R}}, \ \ \ 
\eta_{\mu}^{\dagger} = e^{i\hat{R}} \beta_{\mu}^{\dagger} e^{-i\hat{R}},
\label{eq:unitary2}
\end{equation}
where $\hat{R}$ is a one-body Hermitian operator written in the original q.p.
basis as $\hat{R} = \sum_{\mu\nu} R_{\mu\nu} \beta_{\mu}^{\dagger}\beta_{\nu}$.


\subsubsection{General Invariance of the Density Matrix at $T>0$}
\label{subsubsec-invariance}

We now prove that the one-body density matrix $\rho_{ij}$ of Eq.(\ref{eq:rho}) 
is invariant under a rotation of the q.p. operators. More precisely: if we 
start from a HFB vacuum corresponding to the rotated q.p. operators $\eta$ and 
compute the one-body density matrix by using the Bogoliubov transformation 
$\mathcal{W}'$, the result is the same as if we had started from the HFB vacuum 
of the $\beta$ operators using the initial Bogoliubov transformation 
$\mathcal{W}$.

The statistical trace defining $\rho_{ij}$ can be computed in any arbitrary 
many-body basis $|n\rangle$ of the Fock space. Generically, we thus have
\begin{equation}
\rho_{ij} = \sum_{n} \langle n | \hat{D} c_{j}^{\dagger}c_{i} | n\rangle.
\label{eq:def_rho}
\end{equation}
Let us introduce the new set of q.p. operators $\eta$ obtained by the unitary
transformation of Eq.(\ref{eq:unitary1}). We choose the multi-qp states 
$|n\rangle$ of Eq.(\ref{eq:def_rho}) from the vacuum of the rotated q.p. 
operators, that is,
\begin{equation}
|n\rangle \equiv |\eta_{n}\rangle
= \eta_{1}^{\dagger}\cdots\eta_{n}^{\dagger}|\text{vac}_{\eta}\rangle.
\end{equation}
Then, we introduce the Bogoliubov transformation $\mathcal{W}'$ to express
the $c$ operators as a function of the $\eta$ operators. We find
\begin{multline}
\rho_{ij} =
\sum_{\mu\nu} B_{j\mu}B^{*}_{i\nu} \sum_{n}
\langle \eta_{n} | \hat{D} \eta_{\mu}\eta^{\dagger}_{\nu} | \eta_{n} \rangle \\
+
\sum_{\mu\nu} A^{*}_{j\mu}A_{i\nu} \sum_{n}
\langle \eta_{n} | \hat{D} \eta^{\dagger}_{\mu}\eta_{\nu} | \eta_{n} \rangle,
\end{multline}
We now use the property (\ref{eq:unitary2}) to express the multi-qp states
$|\eta_{n}\rangle$ of the $\eta$ operators as a function of the multi-qp states
$|\beta_{n}\rangle$ of the $\beta$ operators. By definition of the multi-qp
states, we find
\begin{equation}
| \eta_{n} \rangle
= \eta_{1}^{\dagger}\cdots\eta_{n}^{\dagger}|\text{vac}_{\eta}\rangle \\
= e^{i\hat{R}} | \beta_{n} \rangle.
\end{equation}
Above, we have the used the property
$e^{-i\hat{R}}|\text{vac}_{\eta}\rangle = |\text{vac}_{\beta}\rangle$. This 
property is the direct consequence of the definition of the vacuum: it is 
the state such that, for all vectors $v$ of the Fock space and any index $i$, 
$\langle v | \eta_{i}|\text{vac}_{\eta}\rangle = 0$. Defining 
$|w\rangle = e^{-i\hat{R}}|v\rangle$, we find 
$\langle w | \beta_{i}e^{-i\hat{R}}|\text{vac}_{\eta}\rangle = 0$. Hence,
$e^{-i\hat{R}}|\text{vac}_{\eta}\rangle$ is the vacuum for operators 
$\beta_{i}$, since the property is valid for all $|w\rangle$. Since the 
vacuum is unique, we must have 
$e^{-i\hat{R}}|\text{vac}_{\eta}\rangle = |\text{vac}_{\beta}\rangle$. 
In matrix form, we can thus write
\begin{equation}
\rho = B^{*}G'B^{T} + AF'A^{\dagger},
\label{eq:rho_rotated}
\end{equation}
if we note
\begin{equation}
\begin{array}{l}
G': G'_{\nu\mu} =
\displaystyle\sum_{n}\langle \beta_{n} | e^{-i\hat{R}}\hat{D} \eta_{\mu}\eta^{\dagger}_{\nu} e^{i\hat{R}}| \beta_{n} \rangle, \\
F': F'_{\nu\mu} =
\displaystyle\sum_{n}\langle \beta_{n} | e^{-i\hat{R}}\hat{D} \eta^{\dagger}_{\mu}\eta_{\nu} e^{i\hat{R}}| \beta_{n} \rangle.
\end{array}
\end{equation}
By virtue of Eq.(\ref{eq:unitary1}),
\begin{equation}
\begin{array}{l}
G'_{\nu\mu} = \displaystyle
\sum_{\alpha\beta} T^{*}_{\mu\alpha} T_{\nu\beta} \sum_{n}
\langle \beta_{n} | e^{-i\hat{R}}\hat{D} \beta_{\alpha}\beta^{\dagger}_{\beta} e^{i\hat{R}}| \beta_{n} \rangle, \\
F'_{\nu\mu} = \displaystyle
\sum_{\alpha\beta}  T_{\mu\alpha} T^{*}_{\nu\beta} \sum_{n}
\langle \beta_{n} | e^{-i\hat{R}}\hat{D} \beta^{\dagger}_{\alpha}\beta_{\beta} e^{i\hat{R}}| \beta_{n} \rangle.
\end{array}
\end{equation}
In matrix form, this leads to
\begin{equation}
G' = TGT^{\dagger}, \ \ \ F' = T^{*}FT^{T},
\label{eq:rotated_R}
\end{equation}
with
\begin{equation}
\begin{array}{l}
G: G_{\beta\alpha} =
\displaystyle\sum_{n}\langle \beta_{n} | e^{-i\hat{R}}\hat{D} \beta_{\alpha}\beta^{\dagger}_{\beta} e^{i\hat{R}}| \beta_{n} \rangle, \\
F: F_{\beta\alpha} =
\displaystyle\sum_{n}\langle \beta_{n} | e^{-i\hat{R}}\hat{D} \beta^{\dagger}_{\alpha}\beta_{\beta} e^{i\hat{R}}| \beta_{n} \rangle.
\end{array}
\end{equation}
Putting everything back together, we find
\begin{align}
\rho
& =  B^{*}G'B^{T} + AF'A^{\dagger}, \\
& =  V^{*}T^{\dagger}TGT^{\dagger}TV^{T} + UT^{T}T^{*}FT^{T}T^{*}U^{\dagger}, \\
& =  V^{*}GV^{T} + UFU^{\dagger}.
\end{align}
Now, it suffices to notice that
\begin{equation}
G_{\beta\alpha}
= \text{Tr} \left(e^{-i\hat{R}}\hat{D} \beta_{\alpha}\beta^{\dagger}_{\beta} e^{i\hat{R}}\right) \\
= (1-f_{\alpha})\delta_{\alpha\beta},
\end{equation}
by cyclic invariance of the trace. Similarly, we find that
$F_{\beta\alpha} = f_{\alpha}\delta_{\alpha\beta}$, so that
\begin{equation}
\rho_{ij} =
\sum_{\alpha} V_{j\alpha}V^{*}_{i\alpha} (1-f_{\alpha})
+
\sum_{\alpha} U^{*}_{j\alpha}U_{i\alpha} f_{\alpha}.
\end{equation}
This shows that the one-body density matrix in the single-particle basis is 
invariant under a unitary transformation of quasiparticle operators among 
themselves at $T>0$. Equation (\ref{eq:rho_rotated}) gives the expression 
of the density matrix after rotation of the qp; Eq.(\ref{eq:rho}) before 
rotation. Both expressions coincide. Using a similar reasoning, one can show 
that the pairing tensor is also invariant under a unitary transformation. 
This implies that both the generalized density $\mathcal{R}$ and the HFB 
matrix (in the s.p. basis) are invariant, $\mathcal{R}'=\mathcal{R}$ and 
$\mathcal{H}'=\mathcal{H}$. Since the Wick theorem guarantees that there 
is a one-to-one correspondence between $\mathcal{R}$ and the quadratic 
operator $\hat{K}$ defining the density operator, see Sec.\ref{subsec-fthfb}, 
we have also $\hat{K}' = \hat{K}$, and most importantly $\mathcal{H}'=\hat{K}'$. 
In other words, the HFB equations are still obeyed in the rotated q.p. basis. 


\subsubsection{Quantum Localization in the Rotated Quasiparticle Basis}
\label{subsubsec-localization}

Although the generalized density $\mathcal{R}$ in the s.p. basis is invariant 
under a unitary transformation of the qp, its form in the basis of the $\eta$ 
operators (the rotated qp basis) is different from what it is in the basis of 
the $\beta$ operators (the original qp basis). Before rotation, the matrix 
$\tilde{\mathcal{R}}$ of the generalized density in the quasiparticle basis 
is diagonal and we have the usual relations
\begin{equation}
\mathcal{R} =
\mathcal{W}\tilde{\mathcal{R}}\mathcal{W}^{\dagger}, \ \ \ \ 
\tilde{\mathcal{R}} =
\left(\begin{array}{cc}
f & 0 \\ 
0 & 1-f
\end{array}\right).
\end{equation}
Introducing the unitary transformation $\mathcal{T}$, we can write
\begin{equation}
\mathcal{R} =
\mathcal{W}'
\left( \begin{array}{cc} T^{*} & 0 \\ 0 & T\end{array} \right)
\left(\begin{array}{cc}
f & 0 \\ 
0 & 1-f
\end{array}\right)
\left( \begin{array}{cc} T^{T} & 0 \\ 0 & T^{\dagger}\end{array} \right)
\mathcal{W}'^{\dagger}.
\end{equation}
Owing to Eq.(\ref{eq:rotated_R}), we can define the matrix of the generalized 
density in the new q.p. basis of the $\eta$ operators defined by Eq.
(\ref{eq:unitary1}) with
\begin{equation}
\tilde{\mathcal{R}}'
=
\left( \begin{array}{cc} T^{*} & 0 \\ 0 & T\end{array} \right)
\left(\begin{array}{cc}
f & 0 \\ 
0 & 1-f
\end{array}\right)
\left( \begin{array}{cc} T^{T} & 0 \\ 0 & T^{\dagger}\end{array} \right).
\end{equation}
In our case, the unitary transformation is given by definition (23) 
in (I), 
\begin{equation}
T = 
\left(\begin{array}{rc}
 \cos\theta & \sin\theta \\
-\sin\theta & \cos\theta
\end{array}\right).
\end{equation}
A simple calculation yields, for the pair $(\mu,\nu)$ of quasiparticles, 
\begin{multline}
T^{*}\left( \begin{array}{rc}
f_{\mu} & 0 \\
0 & f_{\nu}
\end{array}\right)T^{T}
=
\left( \begin{array}{rc}
f_{\mu} & 0 \\
0 & f_{\nu}
\end{array}\right) \\
-
\sin\theta (f_{\mu}-f_{\nu})
\left( \begin{array}{rc}
\sin\theta & \cos\theta \\
\cos\theta & -\sin\theta
\end{array}\right),
\end{multline}
and a similar expression for the term $1-f$. Therefore, the generalized
density matrix loses its diagonal form up to first order in 
$\Delta f = f_{\mu} - f_{\nu}$. In the rotated qp basis, the generalized 
density thus takes the generic form
\begin{equation}
\tilde{\mathcal{R}}' 
=
\left(\begin{array}{cccccccc}
& f_{k}  & & & & & & \\
& & \ddots & & & & & \\
& & & 
\begin{array}{cc}
F_{\mu\mu} & F_{\mu\nu} \\
F_{\nu\mu} & F_{\nu\nu} 
\end{array} & & & \\
  & & & & & 1-f_{k} & & \\
  & & & & & & \ddots  & \\
  & & & & & & & 
\begin{array}{cc}
G_{\mu\mu} & G_{\mu\nu} \\
G_{\nu\mu} & G_{\nu\nu} 
\end{array}\\
\end{array}
\right).
\end{equation}
It can always  be ordered in the form of a block-diagonal matrix, with 
exactly diagonal terms that contain the statistical occupations of the 
q.p. that have not been rotated, and 2$\times$2 non-diagonal blocks 
corresponding to each pair $(\mu,\nu)$ of rotated q.p.

As a consequence of the non-diagonal form of $\tilde{\mathcal{R}}'$, 
the one-body density matrix cannot be expressed as a simple sum over 
single quasiparticle densities. This implies that the coordinate space 
representation of $\rho$ becomes
\begin{equation}
\rho(\gras{r}\sigma,\gras{r}'\sigma')
=
\sum_{k\in\mathcal{S}} \rho_{k}(\gras{r}\sigma,\gras{r}'\sigma')
+
\sum_{\mu\nu\in\mathcal{P}} \rho'_{\mu\nu}(\gras{r}\sigma,\gras{r}'\sigma'),
\end{equation}
where $\mathcal{S}$ refers to the set of q.p. that are not rotated, and 
$\mathcal{P}$ to the set of q.p. that are rotated. The contribution 
$\rho'_{\mu\nu}(\gras{r}\sigma,\gras{r}'\sigma')$ of the rotated
pair $(\mu,\nu)$ of quasiparticles to the total one-body density (which 
is invariant) is
\begin{multline}
\rho'_{\mu\nu}(\gras{r}\sigma,\gras{r}'\sigma')
= \sum_{ij} \phi_{i}(\gras{r}\sigma)\phi_{j}^{*}(\gras{r}'\sigma')\times \\
\left[
B_{i\mu}^{*}G'_{\mu\mu}B_{j\mu} + A_{i\mu}F'_{\mu\mu}A_{j\mu}^{*}
+
B_{i\mu}^{*}G'_{\mu\nu}B_{j\nu} + A_{i\mu}F'_{\mu\nu}A_{j\nu}^{*}
\right.\\
\left.
+
B_{i\nu}^{*}G'_{\nu\mu}B_{j\mu} + A_{i\nu}F'_{\nu\mu}A_{j\mu}^{*}
+
B_{i\nu}^{*}G'_{\nu\nu}B_{j\nu} + A_{i\nu}F'_{\nu\nu}A_{j\nu}^{*}
\right].
\end{multline}
For rotated q.p., the notion of spatial occupation cannot be captured by 
the quantity $N_{\mu}$ alone. We thus redefine the spatial occupation 
of the rotated pairs $(\mu,\nu)$ and $(\nu,\mu)$ of quasiparticles by
\begin{multline}
N'_{\mu\nu} = 
\sum_{i} \left[ B_{i\mu}^{*}G'_{\mu\mu}B_{i\mu} 
+ A_{i\mu}F'_{\mu\mu}A^{*}_{i\mu} \right]\\
+
\sum_{i} \left[ B_{i\mu}^{*}G'_{\mu\nu}B_{i\nu} 
+ A_{i\mu}F'_{\mu\nu}A^{*}_{i\nu} \right]
\label{eq:occ_rot}
\end{multline}
and
\begin{multline}
N'_{\nu\mu} = 
\sum_{i} \left[ B_{i\nu}^{*}G'_{\nu\nu}B_{i\nu} 
+ A_{i\nu}F'_{\nu\nu}A^{*}_{i\nu} \right]\\
+
\sum_{i} \left[ B_{i\nu}^{*}G'_{\nu\mu}B_{i\mu} 
+ A_{i\nu}F'_{\nu\mu}A^{*}_{i\mu} \right]
\label{eq:occ_rot1}
\end{multline}
Note that $N'_{\mu\nu} \neq N'_{\nu\mu}$. A tedious but straightforward 
calculation shows that $N'_{\mu\nu} + N'_{\nu\mu} = N_{\mu} + N_{\nu}$, 
which is nothing but the consequence of the invariance of the density 
matrix under this rotation. Similarly, the spatial occupations in the 
fragment (1) of the rotated pair $(\mu,\nu)$ and $(\nu,\mu)$ of 
quasiparticles now read
\begin{multline}
N'_{1,\mu\nu} = 
\sum_{ij} \left[ B_{i\mu}^{*}G'_{\mu\mu}B_{j\mu} 
+ A_{i\mu}F'_{\mu\mu}A^{*}_{j\mu} \right]d_{ij}(z_{N})\\
+
\sum_{ij} \left[ B_{i\mu}^{*}G'_{\mu\nu}B_{j\nu} 
+ A_{i\mu}F'_{\mu\nu}A^{*}_{j\nu} \right]d_{ij}(z_{N})
\label{eq:occ_rot2}
\end{multline}
and
\begin{multline}
N'_{1,\nu\mu} = 
\sum_{ij} \left[ B_{i\nu}^{*}G'_{\nu\nu}B_{j\nu} 
+ A_{i\nu}F'_{\nu\nu}A^{*}_{j\nu} \right]d_{ij}(z_{N})\\
+
\sum_{ij} \left[ B_{i\nu}^{*}G'_{\nu\mu}B_{j\mu} 
+ A_{i\nu}F'_{\nu\mu}A^{*}_{j\mu} \right]d_{ij}(z_{N})
\label{eq:occ_rot3}
\end{multline}
This simple extension reflects the fact that the two quasiparticles 
forming the pair are not independent anymore.


\subsubsection{Implementation of the Quantum Localization at $T>0$}

In practice, we construct the fission fragments by scanning both the set 
$\mathcal{S}$ of non-rotated q.p. and the set $\mathcal{P}$ of rotated q.p.:
\begin{itemize}
\item For all q.p. $\mu\in\mathcal{S}$, we compute $N_{\mu}$, $N_{1,\mu}$ 
and $N_{2,\mu}$ according to Eq.(\ref{eq:occ}) and Eq.(\ref{eq:occupation1}); 
the q.p. is assigned to fragment (1) if $N_{1,\mu} \geq 0.5N_{\mu}$, to 
fragment (2) otherwise;
\item Let us note $\mathcal{P}_{\mu} = (\mu,\nu)$ the pair of q.p. $\mu$ and 
$\nu$. We have $\mathcal{P} = \bigcup \mathcal{P}_{\mu}$. For each pair 
$\mathcal{P}_{\mu}$, and for each q.p. $\mu$ in this pair, we compute 
$N'_{\mu\nu}$, $N'_{1,\mu\nu}$ and $N'_{2,\mu\nu}$  according to 
Eqs.(\ref{eq:occ_rot})-(\ref{eq:occ_rot3}); the q.p. $\mu$ is assigned to 
fragment (1) if $N'_{1,\mu\nu} \geq 0.5N'_{\mu\nu}$, to fragment (2) otherwise. 
Note that the procedure must be done separately for the q.p. $\mu$ of and 
the q.p. $\nu$ of the pair, since $N'_{\mu\nu} \neq N'_{\nu\mu}$.
\end{itemize}
The result of this procedure is to partition the whole set of quasiparticles 
into two subsets corresponding to the two fragments. The pseudeodensities in 
the fragment can then be formally written as
\begin{eqnarray}
\rho'^{(\text{f})}(\gras{r}\sigma,\gras{r}'\sigma') 
& = & 
\sum_{\mu\in (f)}\sum_{\nu\in\mathcal{P}_{\mu}}\sum_{ij}
\left[ 
B_{i\mu}^{*}G'_{\mu\nu}B_{j\nu} 
\right. \nonumber\\
& & \left.
+ A_{i\mu}F'_{\mu\nu}A_{j\mu}^{*} 
\right] \phi_{i}(\gras{r}\sigma)\phi_{j}^{*}(\gras{r}'\sigma'), 
\label{eq:rho_f_r_rot}\medskip \\
\kappa'^{(\text{f})}(\gras{r}\sigma,\gras{r}'\sigma') 
& = & 
\sum_{\mu\in (f)}\sum_{\nu\in\mathcal{P}_{\mu}}\sum_{ij}
\left[ 
B_{i\mu}^{*}G'_{\mu\nu}A_{j\nu} 
\right. \nonumber\\
& & \left. 
+ A_{i\mu}F'_{\mu\nu}B_{j\nu}^{*} \right]
\phi_{i}(\gras{r}\sigma)\phi_{j}^{*}(\gras{r}'\sigma').
\label{eq:kappa_f_r_rot}
\end{eqnarray}
These relations allow to extend the calculation of the interaction energy 
between the fragments and the fragment internal energies at $T>0$ using the 
formulas given in (I).

We have implemented the localization method in a new module of the DFT solver 
HFODD \cite{schunck2011}. The rotation of the q.p. is first performed in 
the s.p. space, i.e., the matrices $U$ and $V$ of the Bogoliubov transformation 
are rotated according to Eq.(\ref{eq:unitary_implemented}). Using 
Eq.(\ref{eq:rho}) and Eq.(\ref{eq:rho_rotated}), we have checked that the density 
matrix in the s.p. basis (the deformed HO basis in our case) remains invariant 
after the transformation within numerical precision. 

In HFODD, calculations of the nuclear and Coulomb interaction energy are 
carried out in coordinate space. The matrices of the Bogoliubov transformation 
are first transformed into spinors according to
\begin{equation}
\begin{array}{l}
\displaystyle\varphi_{\mu}^{(1)}(\gras{r}\sigma) = -2\sigma\sum_{i} U^{*}_{i\mu}\phi_{i}^{*}(\gras{r}-\sigma), \medskip\\
\displaystyle\varphi_{\mu}^{(2)}(\gras{r}\sigma) = \sum_{i} V^{*}_{i\mu}\phi_{i}(\gras{r}\sigma),
\label{eq:spinors}
\end{array}
\end{equation}
Note that this transformation depends on a specific phase convention. Local 
densities are then defined in terms of these HFB spinors; see 
\cite{dobaczewski2004} for details. We have then checked that the coordinate 
space representations of the densities, as computed from the HFB spinors 
(\ref{eq:spinors}) are also invariant after the q.p. rotation within 
machine precision. 

\begin{figure}[!ht]
\center
\includegraphics[width=\linewidth]{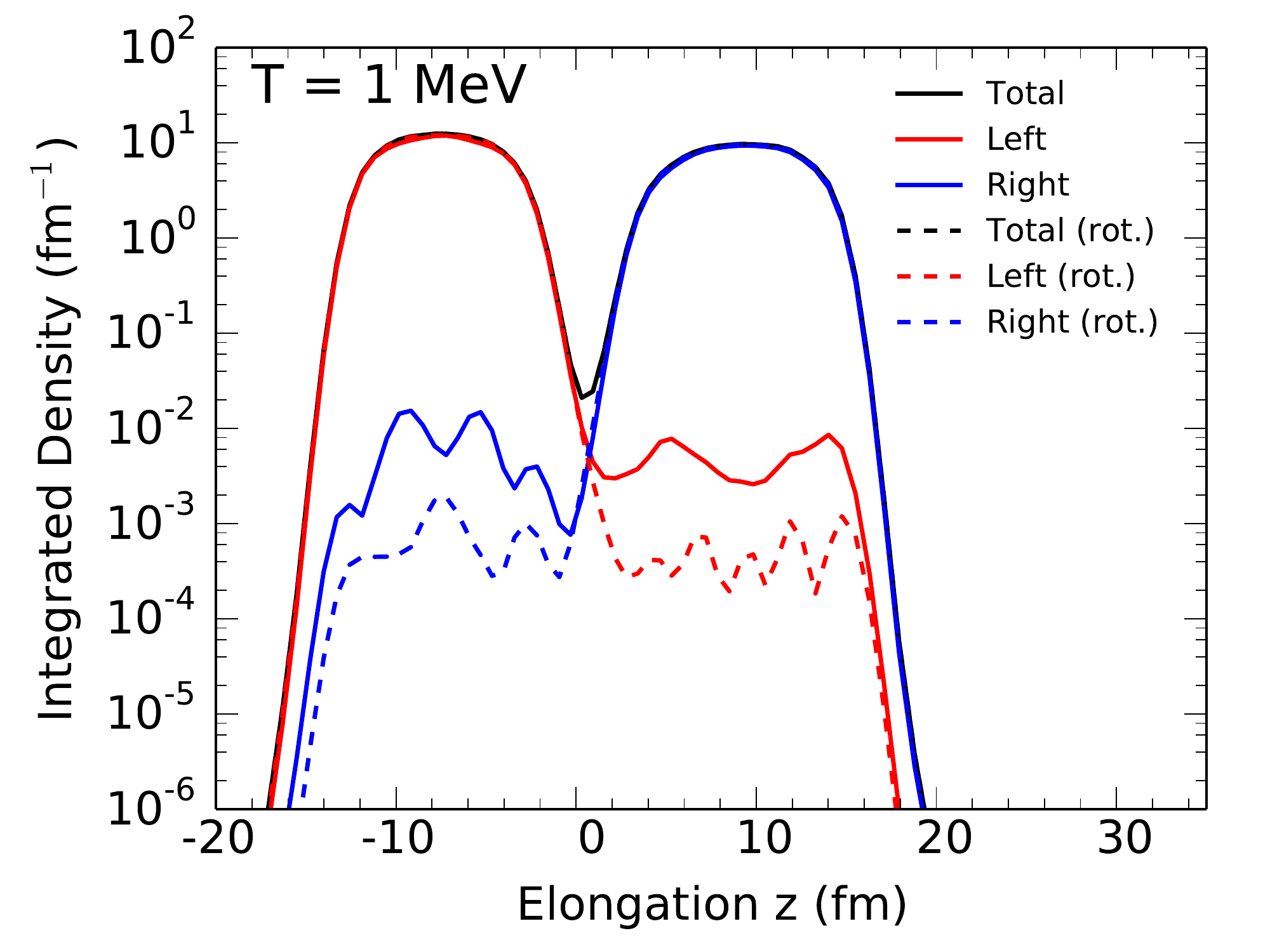}
\caption{(color online) Profile of the integrated nuclear density
$\rho(\gras{z})$ (integrated over $x-$ and $y-$ coordinates) along the
elongation axis $z$ in $^{240}$Pu at $q_{N}=0.7$ and $T=1.0$ MeV before 
(plain lines) and after (dashed lines) rotation of the q.p. wave-functions. }
\label{fig:densityLocalization}
\end{figure}

As an example, figure \ref{fig:densityLocalization} shows the impact of the 
localization on the pseudodensities of the fission fragments. The figure 
shows the one-body density matrix of the compound nucleus before and after 
rotation; it also shows the pseudodensities of the left and right fragment, 
before and after rotation. All densities were integrated along the $x-$ and 
$y-$directions. As at $T=0$, we observe a significant decrease of the tails of 
the densities, of approximately an order of magnitude. The curves labeled 
``Total'' and ``Total (rot.)'', which pertain to the total density before and 
after rotation, are indistinguishable.


\subsection{Coupling to the Continuum}
\label{subsubsec-continuum}

It was demonstrated in Refs.~\cite{dobaczewski1984,dobaczewski1996} based on 
the coordinate space formulation of the HFB equations (in spherical symmetry) 
that the asymptotic conditions for the $(U,V)$ matrices of the Bogoliubov
transformations read
\begin{eqnarray}
U(E,r\sigma) \rightarrow
\left\{
\begin{array}{ll}
\cos(k_{1}r + \delta_{1}) & E > -\lambda\\
e^{-\kappa_{1}r}          & E < -\lambda
\end{array}
\right. \\
V(E,r\sigma) \rightarrow
\left\{
\begin{array}{ll}
\cos(k_{1}r + \delta_{1})  & E < +\lambda\\
e^{-\kappa_{1}r}           & E > +\lambda
\end{array}
\right.
\end{eqnarray}
From these expressions, it was shown that for nuclei with negative Fermi
energy, the local density is always localized, which leads to observables
taking finite values.

At $T>0$, the FT-HFB equations take exactly the same form as at $T=0$, hence
the matrices $U$ and $V$ of the Bogoliubov transformation have the same
asymptotic properties. We summarize in Table~\ref{tab:asymptotic} the localized
or delocalized nature of the $U$ and $V$ matrices of the Bogoliubov
transformation depending on the value of the Fermi level $\lambda$ and the
energy $E$ of the q.p..

\begin{table}[!ht]
\begin{center}
\caption{Localization properties at $T>0$ of the matrices $(U,V)$ of the
Bogoliubov transformation depending on the value of the q.p. energies and the
Fermi level.}
\begin{ruledtabular}
\begin{tabular}{lll}
Fermi Level & q.p. Energy & Localization of $(U,V)$ \\
\hline
\multirow{2}{*}{$\lambda >0 $} & $E > +\lambda$ & $U$ delocalized, $V$ localized \\
                               & $E < +\lambda$ & $U$ delocalized, $V$ delocalized \\
\multirow{2}{*}{$\lambda <0 $} & $E > -\lambda$ & $U$ delocalized, $V$ localized \\
                               & $E < -\lambda$ & $U$ localized,   $V$ localized
\end{tabular}
\label{tab:asymptotic}
\end{ruledtabular}
\end{center}
\end{table}
Contrary to the case at $T=0$, however, the density matrix now takes the form
of Eq.(\ref{eq:rho}), and the additional term $fUU^{\dagger}$ can be
delocalized even for systems with negative Fermi energy. In fact, the set of
quasi-particles can be split into the subset $\mathcal{L}$ of localized,
discrete q.p, with $0\leq E <-\lambda$, and the subset $\mathcal{C}$ of
delocalized, continuous q.p. with $E >-\lambda$. The full density is, of
course, the sum of the two contributions
\begin{equation}
\rho_{ij} = \rho_{ij}^{\text{(loc)}} + \rho_{ij}^{\text{(con)}},
\end{equation}
with
\begin{equation}
\rho_{ij}^{\text{(loc)}} =
\sum_{\mu\in\mathcal{L}}
\left( V^{*}_{i\mu}(1-f)_{\mu}V_{j\mu} + U_{i\mu}f_{\mu}U_{j\mu}^{*} \right),
\end{equation}
and (assuming the continuous spectrum is discretized as, e.g. happens in the
HO basis),
\begin{equation}
\rho_{ij}^{\text{(con)}} =
\sum_{\mu\in\mathcal{C}}
\left( V^{*}_{i\mu}(1-f)_{\mu}V_{j\mu} + U_{i\mu}f_{\mu}U_{j\mu}^{*} \right).
\end{equation}

\begin{figure}[!ht]
\center
\includegraphics[width=\linewidth]{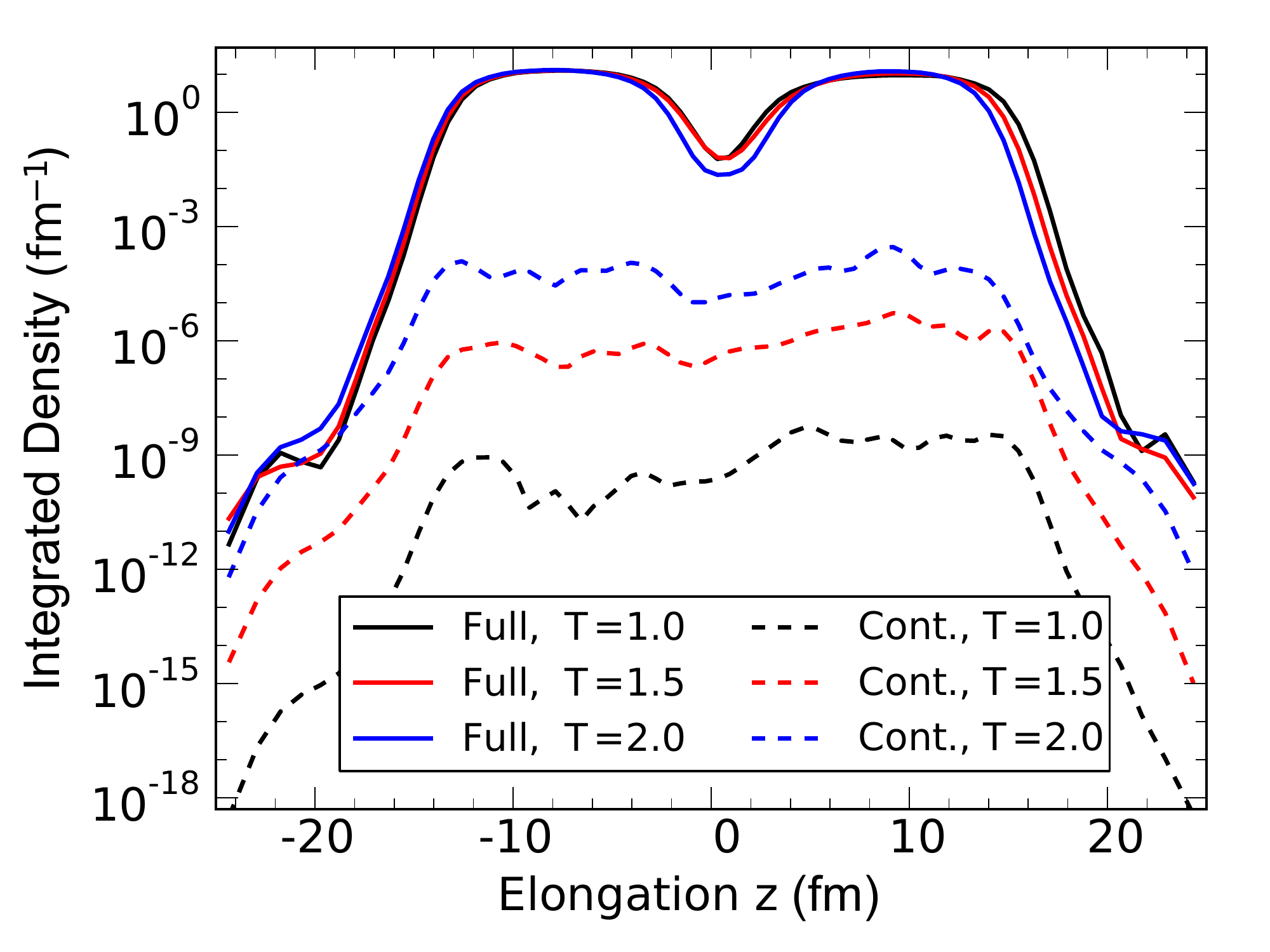}
\caption{(color online) Profile of the isoscalar nuclear density
$\rho_{0}(\gras{r})$ (integrated over $x-$ and $y-$ coordinates) along the
elongation axis $z$ in $^{240}$Pu at $q_{N}=1.0$ and $T=1.0, 1.5, 2.0$ MeV. The
plain lines correspond to the full density, the dashed lines to the term
$fUU^{\dagger}$ only of $\rho^{(\text{con})}$.
}
\label{fig:densityProfiles}
\end{figure}

In Fig.~\ref{fig:densityProfiles}, we illustrate this result by showing the
profile of the total isoscalar density $\rho_{0}(\gras{r})$ along the
elongation axis of $^{240}$Pu in the scission region, at $q_{N}=1.0$, together
with the contribution of the term $fUU^{\dagger}$ to its delocalized
contribution $\rho_{ij}^{\text{(con)}}$. Curves are shown at $T=1.0, 1.5, 2.0$
MeV. At $T=1.0$ MeV, the energy of the incident neutron is of the order of $25$
MeV, while it is more than $70$ MeV at $T=2.0$ MeV. Yet, even at such a high
excitation energy and after integrating over the transverse coordinates $x$ and
$y$, the contribution of the term $fUU^{\dagger}$ to the total density is at
most of the order of $10^{-4}$. We note that the use of the one-center HO basis
induces numerical limitations: the tails of the densities at the boundaries of
the domain are not physical but a consequence of the Gaussian asymptotic
behavior of the basis functions (which is visible as a roughly parabolic
decrease of the density near $z=\pm 20-25$ fm).

\begin{figure}[!ht]
\center
\includegraphics[width=\linewidth]{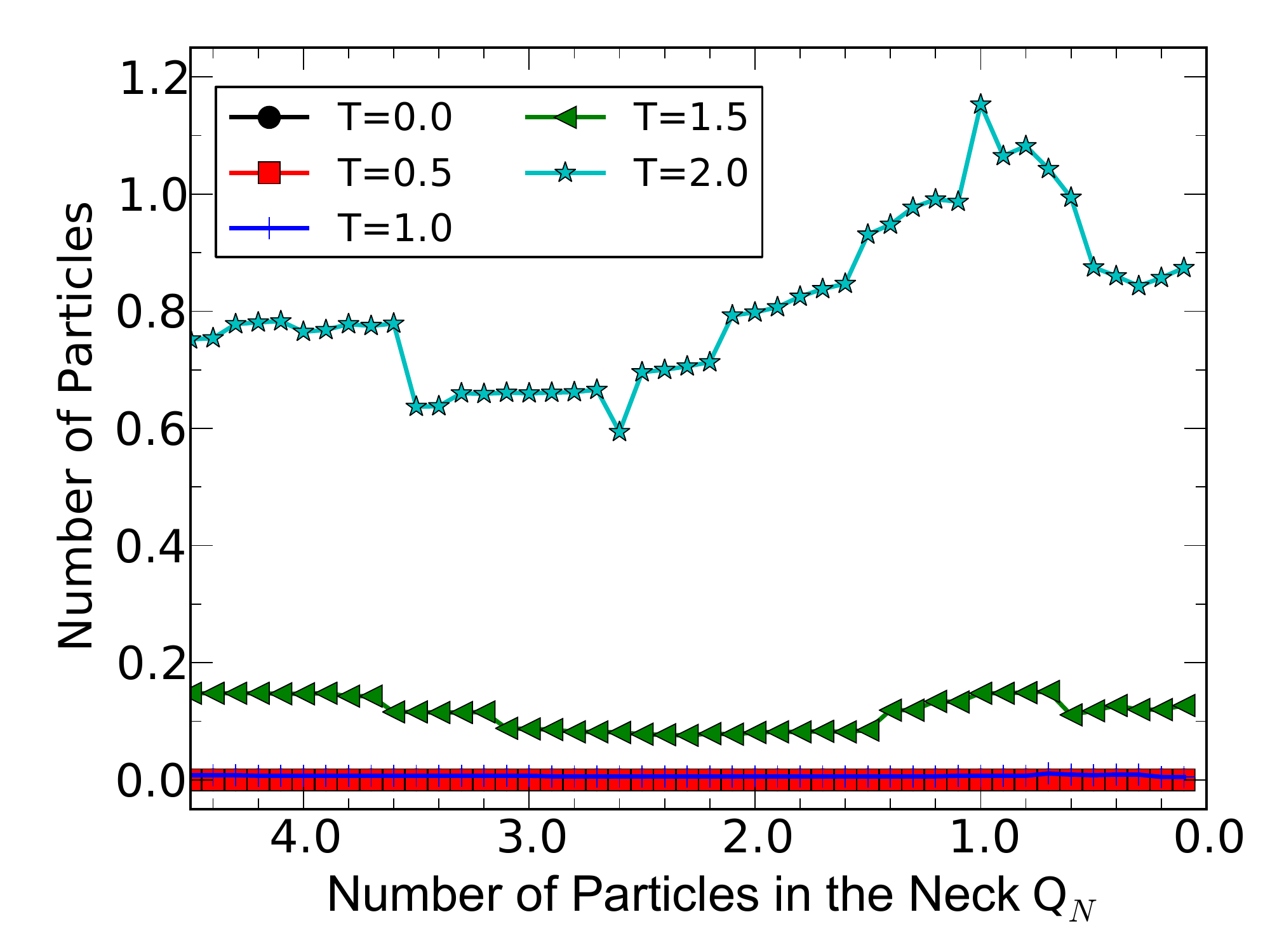}
\caption{(color online) Total number of delocalized quasi-particles as a
function of the number of particles in the neck for various temperatures.
}
\label{fig:particleContinuum}
\end{figure}

The densities can be further integrated over $z$ to provide an estimate of the
number of particles originating from the q.p. located in the continuum. This is
shown in Fig.~\ref{fig:particleContinuum} as a function of $q_{N}$ for five
values of the nuclear temperature, $T=0, 0.5, \dots, 2.0 $ MeV. For
temperatures below $1.5$ MeV, the number of particle is virtually zero; only
beyond 1.5 MeV is the contribution noticeable, with up to about 1 particle in
the continuum at $T=2.0$ MeV. Because of the unphysical spatial truncation of
q.p. wave-functions induced by the asymptotic behavior of the basis functions,
it may be possible that the actual coupling to the continuum is a little
stronger. It is, however, unlikely that the effect is strong enough to have a
sizable impact on the physics of neutron-induced fission.

We have thus shown that, in the regime of temperatures relevant to the
description of induced nuclear fission, the coupling to the continuum remains
essentially negligible. Our results are fully compatible with estimates
published in the literature. Indeed, early works in the context of the
finite-temperature Hartree-Fock theory suggested that the effect of the
continuum becomes significant only at $T > 4$ MeV \cite{bonche1984}. In the
follow-up paper by the same authors, the density of neutron vapor in $^{208}$Pb
was shown to be $0.5 10^{-2}$ fm $^{-3}$ at $T = 7$ MeV \cite{bonche1985}. More
recent estimates obtained at the fully FT-HFB level with a coordinate-space
solver also suggest a total number of particles in the continuum of 0.2 at
$T = 1.5$ MeV in the superheavy element $Z=114$, $N=178$ \cite{pei2009}. These
results are noteworthy, because they justify {\em a posteriori} the validity of
the model of the compound nucleus to describe induced fission.


\subsection{Fragment Interaction Energy and Kinetic Energy}
\label{subsec-fragments}

Based on the JCN analysis presented in Sec.~\ref{subsec-region}, we have
identified the range $q_{N} \in [ 0.1-3.0 ]$ as the scission region, with
$q_{N} \approx 0.2-0.3$ as the most likely scission point (at low
temperatures). Using the generalized quantum localization procedure of
Sec.~{\ref{subsec-localization}, we have computed the fission fragment
interaction energy and total kinetic energy before and after localization for
the range of temperatures $0 \leq T \leq 1.75$ MeV. For $T\geq 1.50$ MeV, the
localization method begins to break down: on the one hand, the number of
possible pairs meeting the criteria for rotation becomes very large and the
procedure becomes very time-consuming; in addition, it does not always succeed
in fully localizing the fragments. This may be an indirect effect of the
coupling to the continuum discussed in the previous section.

\begin{figure}[!ht]
\center
\includegraphics[width=\linewidth]{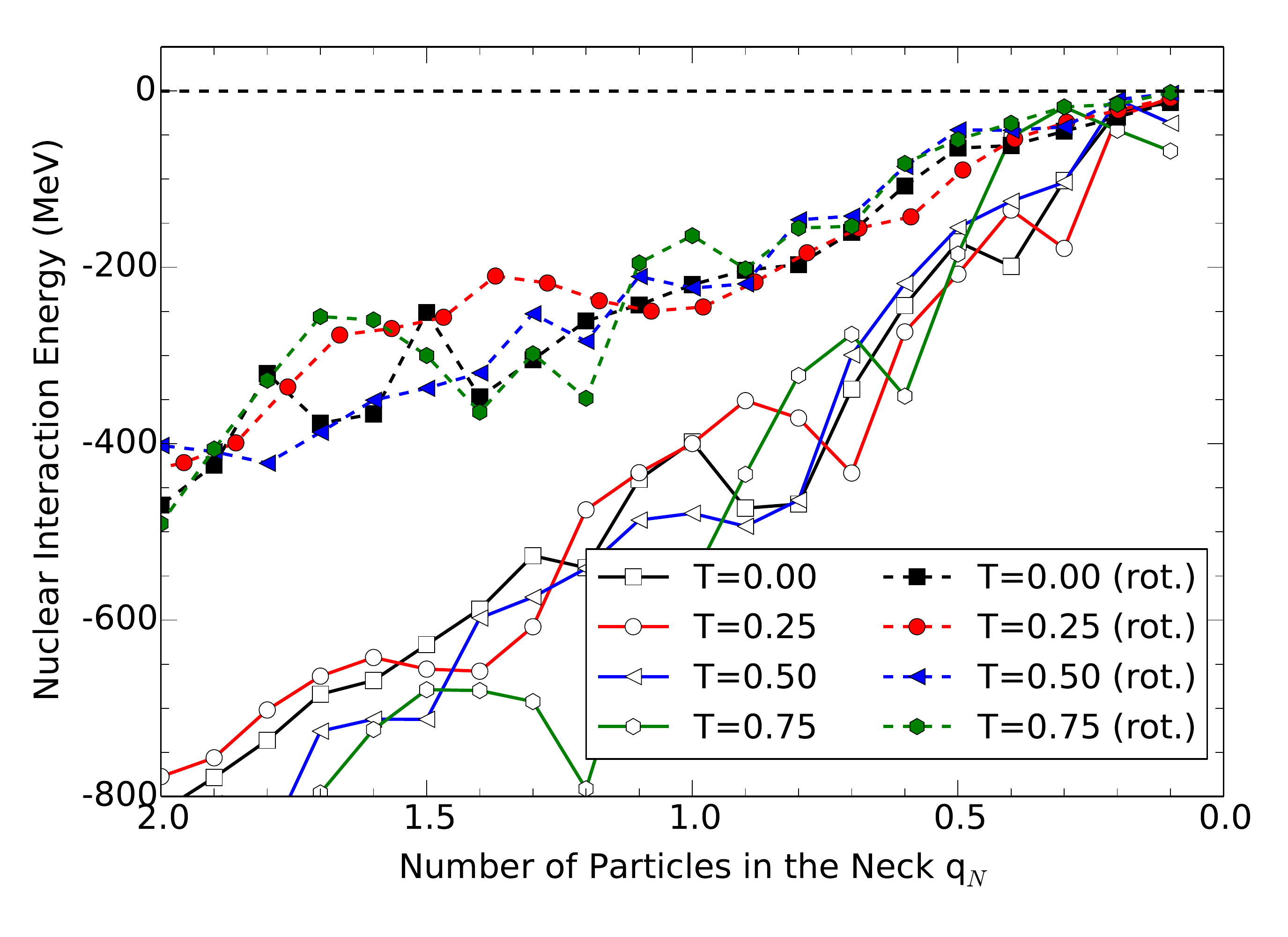}
\caption{(color online) Skyrme interaction energy between the fission fragments
of $^{240}$Pu as a function of the number of particles in the neck for the SkM*
functional at low temperatures $0.00 \leq T \leq 0.75$ MeV. Solid curves with
open symbols correspond to the calculation before the localization is applied,
dashed curves with filled symbols to the localized q.p.
}
\label{fig:interaction_energy_low}
\end{figure}

It is interesting to distinguish two temperature regimes. In the range
$0 \leq T \leq 0.75$ MeV, which is depicted in
Fig.~\ref{fig:interaction_energy_low}, there are relatively few qualitative
differences between the zero-temperature case and the finite-temperature
results: the nuclear interaction energy is of the same order of magnitude at 
all $T$, both before and after quantum localization. This is consistent with 
the earlier observation in Sec.~\ref{subsec-pathway} that the potential energy 
surface does not change dramatically in this temperature range. As in (I), we 
note relatively large fluctuations of the interaction energy as a function of
$q_{N}$, especially before localization. To a large extent, these fluctuations
reflect the binary nature of the partitioning of the nucleus in two (entangled)
fragments: a given q.p. could be assigned to one fragment for a given $q_{N}$
and to the other at $q_{N} + \delta q_{N}$, especially if its localization
$\ell$ indicator is close to 0.5. After localization, such fluctuations are
strongly attenuated but do not disappear entirely, since there remain a few
q.p. that can not be properly localized \cite{younes2011}. In addition, small
discontinuities in the unconstrained collective variables can also contribute
to the fluctuations of interaction energy.

\begin{figure}[!ht]
\center
\includegraphics[width=\linewidth]{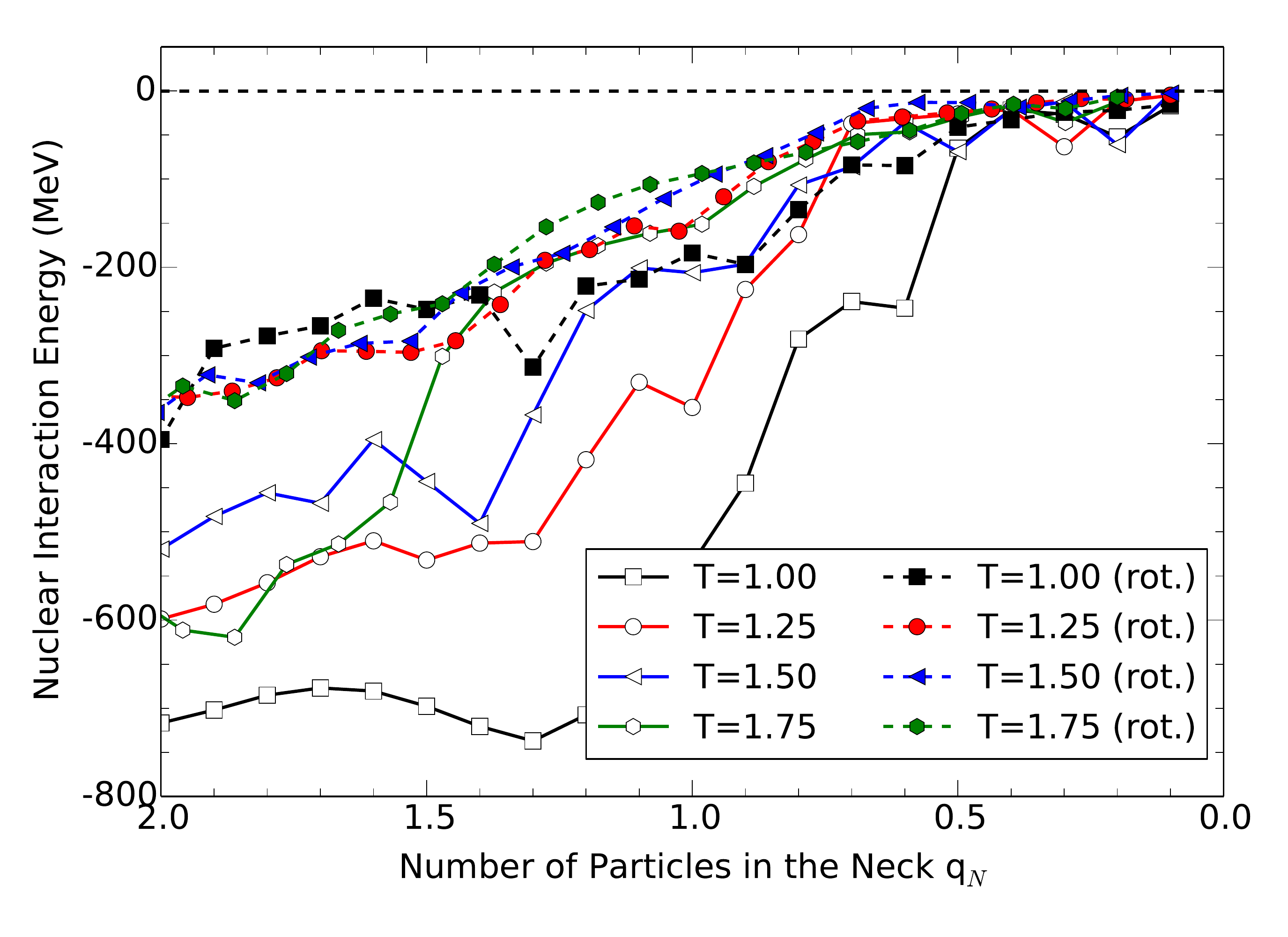}
\caption{(color online) Same as Fig.~\ref{fig:interaction_energy_low} in
the higher temperature regimes $1.00 \leq T \leq 1.75$ MeV.
}
\label{fig:interaction_energy_high}
\end{figure}

In the higher temperature regime, the effect of changes in temperature becomes
more visible. From a purely topological point of view, the scission point as
determined by the JCN is pushed back from $q_{N} \approx 0.2-0.4$ to
$q_{N} \approx 0.5 - 0.9$. This observation is confirmed by the behavior of the
nuclear interaction energy: As a function of $q_{N}$, the interaction energy
goes to zero faster as $T$ increases. This trend is already clearly visible
before localization, the effect of which is to make it more pronounced.
Qualitatively, these results show that the system tends to break with a thicker
neck than at lower temperatures, in a manner somewhat similar to glass.

\begin{figure}[!ht]
\center
\includegraphics[width=\linewidth]{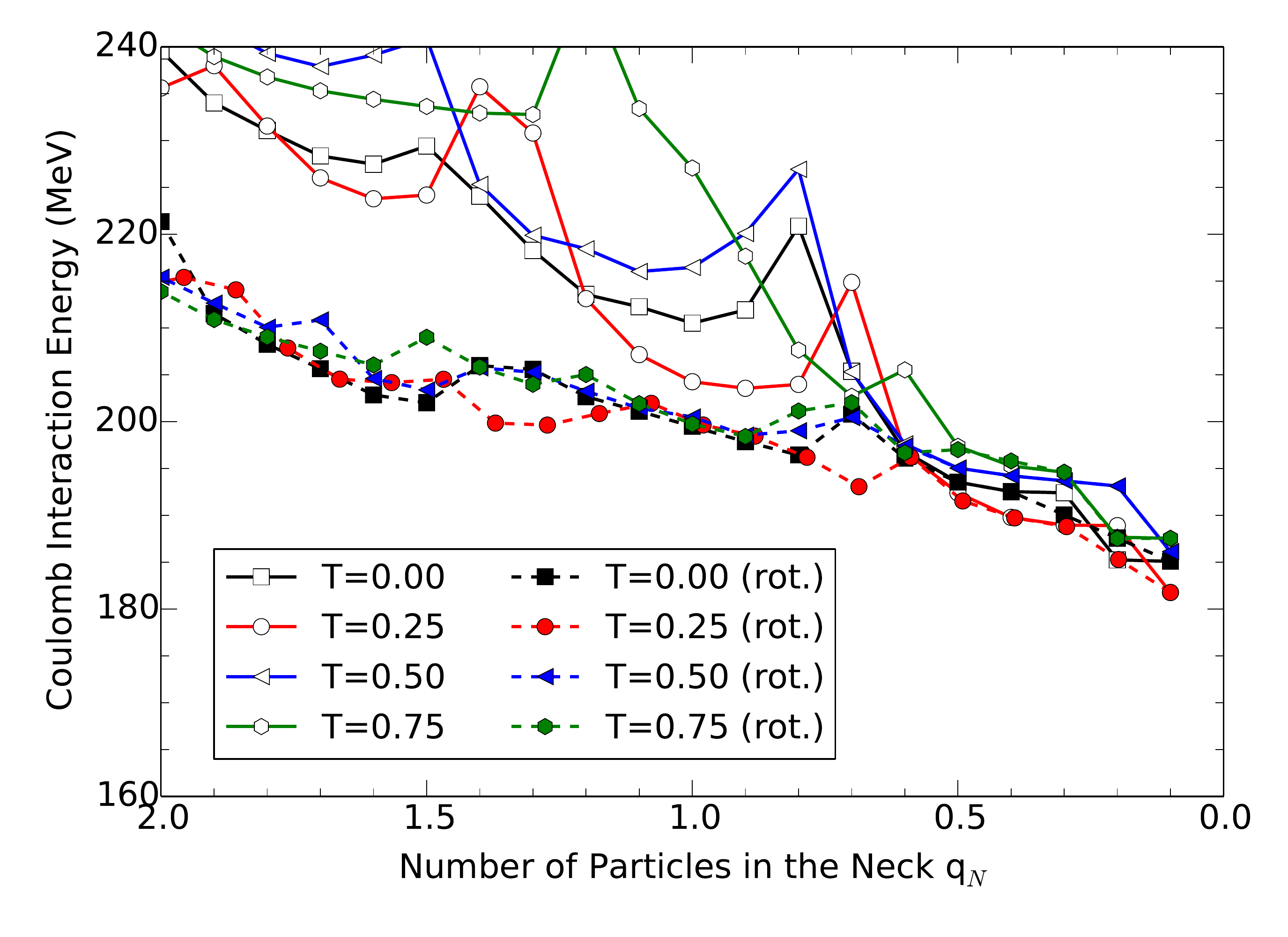}
\caption{(color online) Direct Coulomb interaction energy in the fission of 
$^{240}$Pu as a function of the number of particles in the neck for the SkM* 
functional for temperatures in the range $0.00 \leq T \leq 0.75$ MeV.
}
\label{fig:TKE_low}
\end{figure}

We show in Fig.~\ref{fig:TKE_low} the variations of the direct Coulomb
interaction energy along the $q_{N}$ trajectory at low temperatures, which are
the most relevant to applications of neutron-induced fission. We notice again
the smoothing effect of the localization method, especially at large $q_{N}$
values, where the fragments are still heavily entangled. We also remark that
the effect of the temperature is weak, which is compatible with experimental
evidence, which a variation of about 2 MeV in TKE over a 5 MeV range of
neutron energies \cite{madland2006}. For $q_{N} =0.2$, which the JCN analysis
identifies as the most likely scission configuration, the Coulomb interaction
energy seems first to increase with temperature, from about 185 MeV up to
approximately 195 MeV at $T=0.50$ MeV (corresponding to $E^{*} \approx 8-10 $
MeV excitation energy in the compound nucleus), before decreasing as
temperature keeps on increasing. However, it is clear from the figure that the
amplitude of the energy fluctuations along the $\hat{Q}_{N}$ path in the
scission region are quite large, so these results should be taken with a grain
of salt.

\begin{figure}[!ht]
\center
\includegraphics[width=\linewidth]{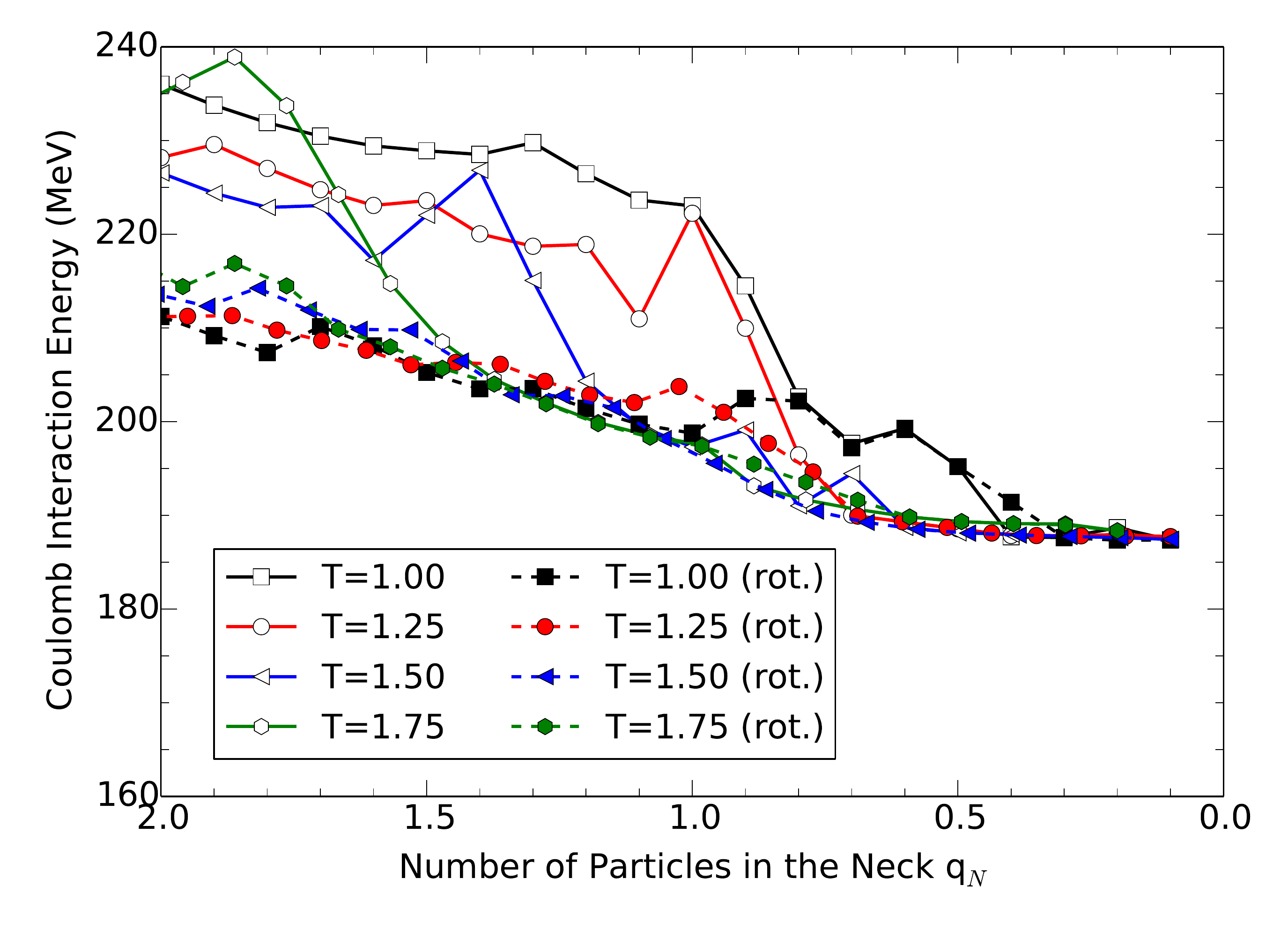}
\caption{(color online) Same as Fig.~\ref{fig:TKE_low} in the higher
temperature range $1.00 \leq T \leq 1.75$ MeV.
}
\label{fig:TKE_high}
\end{figure}

Finally, figure \ref{fig:TKE_high} shows the evolution of the direct Coulomb 
interaction energy along the $\hat{Q}_{N}$ path at higher temperatures 
$T\geq 1.00$ MeV. This corresponds to incident neutron energies larger than 
25 MeV. In this regime, pairing correlations have vanished entirely. Quite 
surprisingly, the total Coulomb interaction energy is nearly constant at low 
$q_{N}$ values, and this constant value is the same for all temperatures. 
Considering the large uncertainties of the current calculations, it is 
premature to draw definitive conclusions, but this point calls for further 
studies.

The calculations presented here are clearly schematic and have yet to reach 
the accuracy obtained from evaluations \cite{madland2006}. We recall that 
the goal of this paper is to set up a framework based on finite-temperature 
DFT that can be used in more systematic studies. In particular, it becomes 
clear from figures \ref{fig:interaction_energy_low}-\ref{fig:TKE_high} that
scission configurations must be identified from a PES that is fully continuous, 
which should remove some of the fluctuations observed here. This can be 
achieved by considering simultaneously all relevant collective variables, i.e, 
at least $\hat{Q}_{20}$, $\hat{Q}_{22}$, $\hat{Q}_{30}$, $\hat{Q}_{40}$ and
$\hat{Q}_{N}$, together with the temperature. In order to compare theoretical
predictions with experimental data, which is based on the average total kinetic
energy, the local enlargment of the collective space should be repeated for all
fragmentations observed in the $^{239}$Pu(n,f) reaction. Improvements on the
quantum localization methods are also possible. Work along these lines is
currently under way.


\section{Conclusions}
\label{sec-conclusions}

One of the main challenges for a theory of induced fission is the need to
accurately describe the (possibly high) excitation energy of the compound
nucleus. In this paper, we have adopted the finite-temperature nuclear density
formalism to describe neutron-induced fission:
\begin{itemize}
\item We have validated the nuclear DFT framework at finite-temperature for the
description of induced fission. In particular, we have given a prescription to
relate the excitation energy of the compound nucleus to the nuclear temperature
of the FT-HFB theory. Following Ref.~\cite{pei2009}, we have confirmed the
validity of the Maxwell relations of thermodynamics over the entire fission
pathway, with the exception of the scission region (unless there are enough
collective variables to make the potential energy surface continuous).
\item We have quantified the effect of the incident neutron energy on the
fission barriers of the compound nucleus $^{240}$Pu. In particular, we have
found that fission barriers slightly increase in the energy range 
$E_{n} = 0 - 5$ MeV; at higher neutron energies, the trend is reversed and 
fission barriers decrease monotonically. We stress that, in the energy range 
of interest in applications of induced fission, ($E_{n} = 0 - 14$ MeV), the 
barriers are lower by at most 15\%. While this can have a significant impact 
on fission observables, in particular fission probabilities, the effect is not 
as dramatic as may have been expected from, e.g., studies of cold fusion in 
superheavy nuclei \cite{sheikh2009}.
\item We have given a microscopic foundation at $T>0$ of the central hypothesis
of induced fission as a two-step process based on the decay of a compound
nucleus. Indeed, we have confirmed that the coupling to the continuum induced
by the finite temperature is negligible at least up to 50 MeV of excitation
energy ($T \approx 1.5$ MeV) and remain small even at larger excitation
energies.
\item We have generalized the quantum localization method of
Ref.~\cite{younes2011} to the case of the finite-temperature DFT, showing that the
method remains applicable up to $T \approx 1.5$ MeV. We have found that
scission tends to occur at larger values of the number of particles in the neck
as temperature increases.
\end{itemize}

In principle, the finite-temperature DFT framework should allow us to compute
the excitation energy of the fragment in a fully microscopic way. There are,
however, multiple caveats. First of all, we have seen that the position of the
scission point changes with temperature. The charge and mass of the fission
fragments also change: the evolution of a given fragment $(Z,N)$ as a function
of the excitation energy of the compound nucleus can not be obtained from a
single fission pathway only, but requires the full local scission
hyper-surface. An additional difficulty is that both the charge and mass of the
fragments are non-integer numbers, through both quantum and statistical
fluctuations. Of course, we may perform HFB calculations for the fragments by
imposing that $\langle \hat{Z}\rangle$ and $\langle \hat{N}\rangle$ take any
value, including fractional ones, but it is not clear how accurate this
approximation would be.

In this work, we have restricted ourselves to a static view of the fission
process. A dynamical treatment of the process would certainly require an
extension of the microscopic theory of collective inertia at finite
temperature. This would allow both fully consistent computations of spontaneous
fission half-lives in the commonly adopted WKB approximation and calculations
of fission yields and energy distributions in the time dependent generator
coordinate method such as in Refs.~\cite{goutte2005,goutte2004,berger1989}.


\bigskip
\begin{acknowledgments}
Stimulating discussions with W. Younes, D. Gogny, D. Regnier, and J. Randrup 
are very gratefully acknowledged. We are also thankful to W. Nazarewicz and 
J.C. Pei for useful comments. This work was partly performed under the auspices 
of the U.S.\ Department of Energy by Lawrence Livermore National Laboratory 
under Contract DE-AC52-07NA27344. Funding was also provided by the U.S.\ 
Department of Energy Office of Science, Nuclear Physics Program pursuant to 
Contract DE-AC52-07NA27344 Clause B-9999, Clause H-9999, and the American 
Recovery and Reinvestment Act, Pub. L. 111-5. Computational resources were 
provided through an INCITE award ``Computational Nuclear Structure'' by the 
National Center for Computational Sciences (NCCS) and National Institute for 
Computational Sciences (NICS) at Oak Ridge National Laboratory, and through 
an award by the Livermore Computing Resource Center at Lawrence Livermore 
National Laboratory. Thanks are also due to the UK Engineering and Physical 
Sciences Research Council, under Grant EP/J013072/1.
\end{acknowledgments}


\bibliographystyle{apsrev4-1}
\bibliography{temperature}

\end{document}